\documentclass[superscriptaddress,twocolumn,prb,aps,showpacs]{revtex4-1}

\usepackage{latexsym}
\usepackage{graphicx}
\usepackage{hyperref}
\usepackage{graphics,graphpap,amssymb}
\usepackage{amsmath}
\usepackage{parskip}
\usepackage{epsfig}
\usepackage{setspace}
\usepackage{setspace}
\usepackage{subfigure}
\usepackage{relsize}
\usepackage{float}

\newcommand{\re}{\operatorname{Re}}

\newcommand{\im}{\operatorname{Im}}

\begin{document}
\title{Non-Hermitian Localization in Biological Networks}




\author{Ariel Amir}
\affiliation {School of Engineering and Applied Sciences, Harvard University, Cambridge, MA 02138, USA}

\author{Naomichi Hatano}
\affiliation {Institute of Industrial Science, University of Tokyo, Komaba, Meguro, Tokyo 153-8505, Japan}

\author{David R. Nelson}
\affiliation {School of Engineering and Applied Sciences, Harvard University, Cambridge, MA 02138, USA}
\affiliation {Department of Physics, Harvard University, Cambridge, MA 02138, USA}

\begin {abstract}
We explore the spectra and localization properties of the $N$-site
banded one-dimensional non-Hermitian random matrices that arise
naturally in sparse neural networks. Approximately equal numbers of
random excitatory and inhibitory connections lead to spatially
localized eigenfunctions, and an intricate eigenvalue spectrum in the
complex plane that controls the spontaneous activity and induced
response. A finite fraction of the eigenvalues condense onto the real
or imaginary axes. For large $N$, the spectrum has remarkable symmetries
not only with respect to reflections across the real and imaginary
axes, but also with respect to $90^\circ$ rotations, with an unusual
anisotropic divergence in the localization length near the origin.
When chains with periodic boundary conditions become directed, with a
systematic directional bias superimposed on the randomness, a hole
centered on the origin opens up in the density-of-states in the
complex plane. All states are extended on the rim of this hole, while
the localized eigenvalues outside the hole are unchanged. The
bias dependent shape of this hole tracks the bias independent contours of constant
localization length.  We treat the large-$N$ limit by a combination of
direct numerical diagonalization and using transfer matrices, an
approach that allows us to exploit an electrostatic analogy connecting
the ``charges" embodied in the eigenvalue distribution with the
contours of constant localization length. We show that similar results
are obtained for more realistic neural networks that obey ``Dale's Law"
(each site is purely excitatory or inhibitory), and conclude with
perturbation theory results that describe the limit of large bias g,
when all states are extended. Related problems arise in random
ecological networks and in chains of artificial cells with randomly
coupled gene expression patterns.
\end {abstract}
\maketitle

\section{Introduction}
The simplest models of neural networks assume long-range connectivity between individual neurons in the brain, leading to synaptic matrices $\boldsymbol M(i,j)$  with connection strengths approximately independent of the separation $r_{ij} =\left|\vec{r}_{i} -\vec{r}_{j} \right|$ in three dimensions.  The eigenvalue spectrum of $\boldsymbol M(i,j)$ controls the spontaneous activity and induced response of the network, and much is known when its elements are chosen from simple random matrix ensembles in the limit of large matrix rank \textit{N.}   For example, classic treatments of the spectra of real symmetric random matrices leading to the Wigner semi-circular density-of-states describing the distribution of real eigenvalues~\cite{mehta2004random,brody1981random} have been generalized by Sommers \textit{et al}.~\cite{sommers1988spectrum} to allow for a tunable asymmetry in Gaussian probability distributions for the matrix elements $\boldsymbol  M(i,j)$ and $\boldsymbol M(j,i)$.   These authors introduce a parameter that interpolates between the Hermitian limit ($\boldsymbol  M(i,j)= \boldsymbol M(j,i)$) studied by Wigner~\cite{wigner}, and the case of fully asymmetric matrices where $\boldsymbol M(i,j)$ and $\boldsymbol M(j,i)$ are independent random variables. In the latter, non-Hermitian limit, the semi-circular eigenvalue distribution on the real axis is replaced by the ``Circular Law''~\cite{ginibre,girko1984circular}, where the eigenvalues are now uniformly distributed inside a circle in the complex plane, with a vanishing fraction lying outside the circle in the limit $N \rightarrow \infty$.   For the general case, Sommers \textit{et al}.\ found that the eigenvalue distribution is uniform inside an \emph{ellipse}, whose aspect ratio along the real and imaginary axes varies with the amount of non-Hermiticity~\cite{sommers1988spectrum}.

As pointed out by Rajan and Abbott~\cite{abbott}, typical applications to neuroscience require that each node in a synaptic conductivity network be either purely excitatory or inhibitory (Dale's Law), which leads to constraints on the signs of the matrix elements $\boldsymbol M(i,j)$:  all entries in a row describing  an excitatory neuron must be positive or zero, and all entries in an inhibitory row must be negative or zero.    These authors then studied eigenvalue spectra of random matrices with long range connectivity, with excitatory and inhibitory networks drawn from distributions with different means and with equal or different standard deviations.   When the strengths of the excitatory and inhibitory connections are appropriately balanced, with equal standard deviations, the eigenvalue distributions can be made to obey the Circular Law by imposing a mild constraint. However, when the standard deviations are different, the eigenvalue density becomes non-uniform within a circle in the complex plane.

Less is known for \emph{N}-site \textit{banded} random matrices with signed matrix elements, which might be an approximate model for neural networks such that $\boldsymbol M(i,j)\sim \exp [-|\vec{r}_{i} -\vec{r}_{j} |/l]$, where $l^{d} $ (in d-dimensions) is often a large volume containing as many as $10^{5} $ neurons.  On spatial scales larger than $l$, the synaptic connectivity matrix becomes sparse, with the largest elements concentrated along the diagonal. Banded Hermitian random matrices in $d$ dimensions, frequently studied in the context of solid-state physics, have long been known to have eigenvalue spectra characterized by a large number of spatially \textit{localized} eigenfunctions~\cite{anderson, efros}, and it is this phenomenon that we wish to study here.   To focus on an extreme example of bandedness, consider a matrix describing a one-dimensional chain of $N$ sites, where only the elements $\boldsymbol M(j,j)$, $\boldsymbol M(j,j+1)$ and $\boldsymbol M(j+1,j)$ describing on-site and nearest-neighbor couplings can be nonzero.  If we wish to impose periodic boundary conditions, we will set $\boldsymbol M(i,j)=\boldsymbol M(i,j+N)=\boldsymbol M(i+N,j),\forall i,j$.   If the lattice spacing $a\approx l$, this model is a rough approximation to the denser neural networks discussed above, coarse-grained out to a scale of order $l$, with each site representing the spatially averaged firing rates of many actual neurons. We concentrate here on off-diagonal randomness, and assume that all $\boldsymbol M(j,j)$ are identical, and describe, say, a site-independent damping to a background firing rate. For the Hermitian case, with $\boldsymbol M(j,j+1)=\boldsymbol M(j+1,j)$ chosen from some probability distribution, nearly all states are localized in the limit of large $N$, with the longest localization lengths occurring near the band center and the shortest localization lengths near the band edges~\cite{efros}. See Appendix A for a brief review and numerical illustration of this solid-state physics example, which provides a useful benchmark for the more intricate problem with complex eigenvalues we study here. Chaudhari \textit{et al}.~\cite{chaudhuri2014diversity} have studied a related problem, with Hermitian coupling strengths falling off exponentially in space and random self-couplings (diagonal randomness), in the context of one-dimensional neural networks, as well as localization of the eigenmodes in a non-Hermitian matrix arising not from disorder but from a slow gradient in the diagonal elements.
Here, we study sparse \emph{non-Hermitian} matrices and the localization properties of their eigenmodes. An important feature of our model is the underlying \emph{spatial} structure (the connections are between nearest neighbors in real space), which distinguishes our work from recent, interesting studies of sparse non-Hermitian matrices without such structure~\cite{sparse1,sparse2, sparse4}.


\subsection{From neural networks to random matrices}
\label{sec2}
As stated above, we focus here on \textit{off-diagonal} randomness in the neural connections, which is both non-Hermitian ($\boldsymbol M(j,j+1)\ne \boldsymbol M(j+1,j)$) and, importantly, also allows for $\boldsymbol M(j,j+1)$ and $\boldsymbol M(j+1,j)$ to be of \textit{opposite sign} roughly 50\% of the time. We thus model a set of approximately balanced excitatory and inhibitory nearest-neighbor neural connections in one dimension, and study the localization properties of the intricate complex eigenvalue spectrum that results.  To put our investigations in context, consider first (using a convenient Dirac notation ${\left| j \right\rangle} $ to describe a neuron at site $j$) the spectrum of a simple \textit{Hermitian} 1d tridiagonal matrix with random connections, namely  $ $
\begin{align} \label{GrindEQ__1_}
 {\boldsymbol M}&=-\sum _{j=1}^{N}\left[ s_{j}^{+} {\left| j \right\rangle} {\left\langle j+1 \right|}
 + s_{j}^{-} {\left| j+1 \right\rangle} {\left\langle j \right|} \right] ,  \nonumber\\
 {s}_{j}^{+} &=s_{j}^{-} =s_{0} +s_{j} >0, \nonumber\\
 s_{j} &\in [-\Delta ,\Delta ],\quad \Delta =0.5s_{0}
\end{align}
Here, all eigenvalues are real, and the symmetrical connections ${ s}_{j}^{+}=s_{j}^{-} $ between neighboring sites are guaranteed to be positive by our choice of a relatively narrow ($\Delta<s_{0}$) box distribution for the bond-to-bond fluctuations in the connection strengths relative to the background level $s_{0} $, and we have subtracted off a diagonal contribution, assumed to be site-independent.  As shown in Appendix A, the localization length $\xi (\varepsilon )$ of the eigenfunctions diverges near the band center at energy $\varepsilon=0$. The quantity $\xi (\varepsilon )$ describes the spatial scale over which an eigenfunction with energy $\varepsilon$ is nonzero.   If the eigenfunction $\phi _{\varepsilon } (j)$ is large near a ``center of localization'' $j^*$, then roughly speaking its envelope decays like $\exp [-|j-j^*|/\xi (\varepsilon )]$. The localization length $\xi (\varepsilon )$ is known to diverge logarithmically~\cite{ziman}, $\xi (\varepsilon )\sim \log (1/\varepsilon ^{2} )$, as $\varepsilon \to 0$. As discussed in Appendix A, for one-dimensional Hermitian localization problems there is an elegant relation connecting the density-of-states $\rho (\varepsilon )$ to the localization length $\xi (\varepsilon )$, known as the Thouless relation~\cite{thouless}. In this case, the Thouless relation implies a strongly diverging density-of-states, $\rho (\varepsilon )\sim 1/[\left|\varepsilon \right|\log ^{3} (1/\varepsilon ^{2} )]$, near the origin. We shall see echoes of these results later in this paper.

\begin{figure}[t]
\centering
\includegraphics[width=0.45\textwidth]{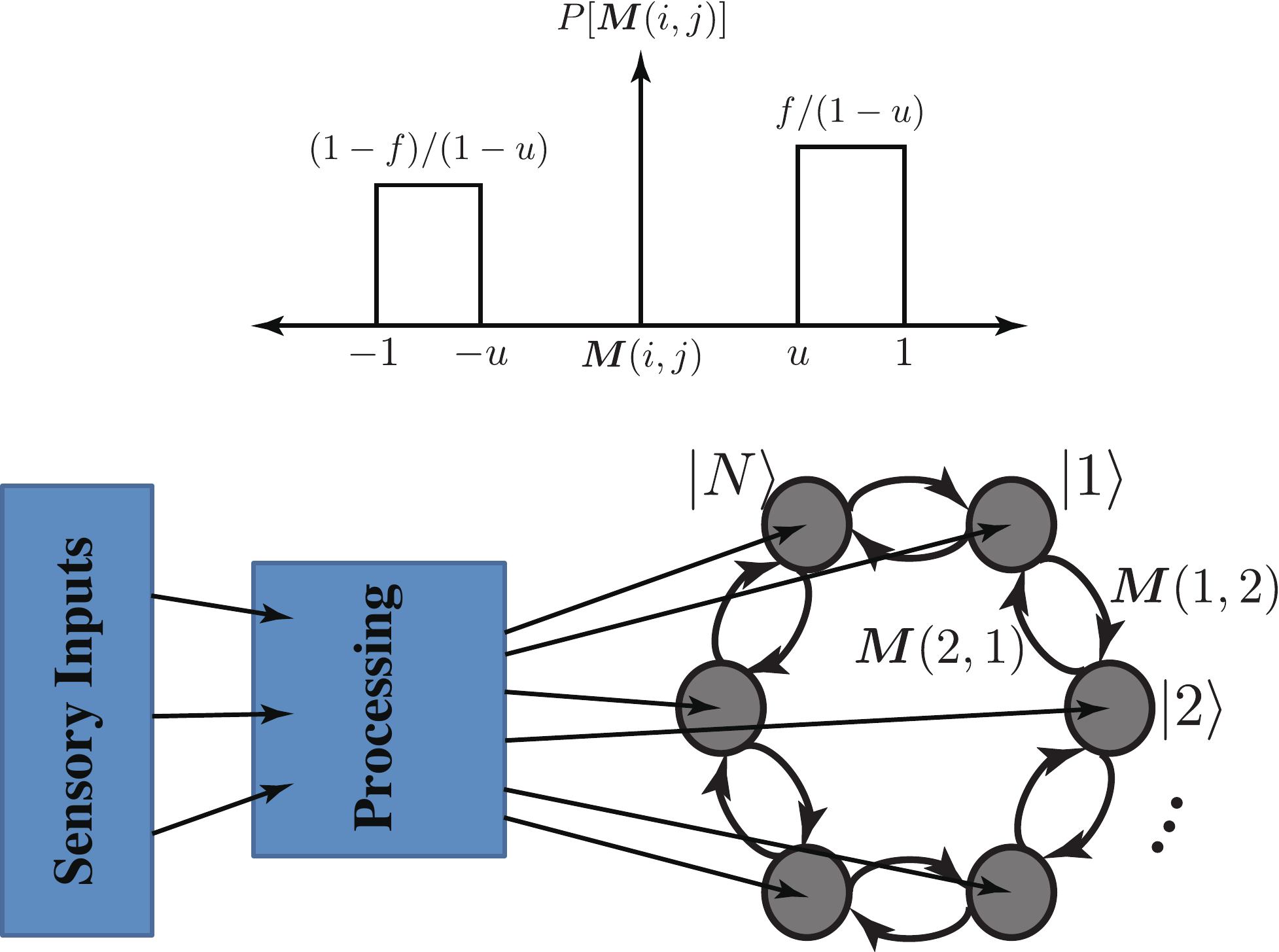}
\caption{Schematic of a 1d neural network problem with periodic boundary conditions.   Sensory inputs, possibly after a processing step, are sent via feed-forward couplings into a circular ring of $N$ neurons $|j\rangle, j = 1,... N$, with nearest-neighbor excitatory and inhibitory connections.  $\boldsymbol M(1,2)$ and $\boldsymbol M(2,1)$ can not only be unequal, but also of opposite sign, if one direction is excitatory and the other inhibitory. Inset shows the probability distribution of a generic nearest-neighbor coupling strength $s$.}
\label{fig1}
\end{figure}

We study here a generalization of Eq.~\eqref{GrindEQ__1_} that arises in one-dimensional neural networks with random excitatory and inhibitory nearest-neighbor connections.  Following Chapter 7 of Ref.~[\onlinecite{dayan2001theoretical}], consider the sparse ``recurrent neural network'' shown in Fig.~\ref{fig1}, a chain of nodes with asymmetric connections between nearest neighbors and with periodic boundary conditions.   Sensory inputs, possibly after a processing step, are sent via feed-forward couplings into a circular ring of $N$ neurons ${\left| j \right\rangle} ,j=1,...,N$.   The nearest-neighbor excitatory and inhibitory couplings $\boldsymbol M(j,j+1)$ and $\boldsymbol M(j+1,j)$ can not only be unequal, but also of opposite sign, if one direction is excitatory and the other inhibitory. Consider a model where the average firing rates $v_{i} $ and $v_{j} $ in neurons $i$ and $j$ (a coarse-grained description of the temporal density of discrete spikes in these neurons) are coupled together, and obey
\begin{align} \label{GrindEQ__2_}
\tau \frac{dv_{i} }{dt} =-v_{i} +F(\boldsymbol M_{ij} v_{j} +h_{i} ).
\end{align}
Here $\boldsymbol M_{ij} \equiv \boldsymbol M(i,j)$, $\tau$ is a characteristic neuron time constant (assumed for simplicity to be the same for all neurons) and we use the summation convention.  Inputs to an animal brain from the outside, due to whiskers, retinal cells, olfaction, \textit{etc}.\ (after a possible processing step), are represented by $h_{i} =\boldsymbol W_{ij} u_{j} $,   where the connection matrix $\boldsymbol W_{ij} \equiv \boldsymbol W(i,j)$ and the input firing rates  $\{ u_{i} \} $ represent the feed-forward part of this network. The activation function $F(w)$ (often taken to have a nonlinear sigmoidal shape~\cite{dayan2001theoretical,hertz1991introduction}) insures that the firing rates are bounded above (when inhibitory connections are present, additional constraints can be imposed to insure that the firing rates can never be negative~\cite{dayan2001theoretical}).   Here we assume for simplicity that the activation function is the same for both excitatory and inhibitory connections.
The eigenvalues and eigenvectors of the matrix $\boldsymbol M_{ij}$ are clearly important for understanding the behavior of a linearized version of Eq.~\eqref{GrindEQ__2_},
\begin{align} \label{GrindEQ__3_}
\tau \frac{dv_{i} }{dt} =-v_{i} +\boldsymbol M_{ij} v_{j} +h_{i} ,
\end{align}
where we assume without loss of generality that $F(0)=0$ and $F'(0)=1$.  This linear recurrent network is capable of both selective amplification and input integration~\cite{dayan2001theoretical}. More generally, knowledge of the eigenvalues and eigenfunctions of $\boldsymbol M_{ij}$ is useful for studying spontaneous activity and evoked responses~\cite{shriki2003rate,vogels2005neural}.  Spontaneous activity depends on whether the real parts of any of the eigenvalues are large enough to destabilize the silent state in a linear analysis, and the spectrum of eigenvalues with large real parts provides valuable information about the spontaneous activity in the full, nonlinear models, and about the localization volume determining the size of the active clusters carrying out computations.  Moreover, similar matrices arise when nonlinear problems are linearized about a steady state.

To see why   \textit{random}  neural connections might be relevant, note that these can be formed during development, with many random attachments of axons and dendrites to other neurons.  Then, over time, pruning (loss of connections) and adaptation (strengthening and weakening of various excitatory and inhibitory connections) occur as neural circuits ``learn'' various functions.  The likely result is a mixture of structured and random components.  The spectra and eigenfunctions of completely random sparse neural network chains, with a mixture of inhibitory and excitatory connections, could provide a description of neural activity during the early stages of development, and is also a useful reference model.    Similar justifications have been advanced for studying the dense neural networks that obey Dale's law treated in Ref.~[\onlinecite{abbott}].

\subsection {Model and density-of-states}

 With this motivation, we now discuss the spectra of non-Hermitian matrices that generalize Eq.~\eqref{GrindEQ__1_}, namely
\begin{align} \label{GrindEQ__4_}
{\boldsymbol M}=\sum _{j=1}^{N}\left[s_{j}^{+} e^{g} {\left| j+1 \right\rangle} {\left\langle j \right|} {\rm \; +\; s}_{j}^{-} e^{-g} {\left| j \right\rangle} {\left\langle j+1 \right|} \right],
\end{align}
where for most of this paper we impose periodic boundary connections, ${\left| j+N \right\rangle} ={\left| j \right\rangle}$.   The constant diagonal contribution associated with Eq.~\eqref{GrindEQ__3_} has again been subtracted off.  The connection strengths $s_{j}^{+} $and $s_{j}^{-} $ are independent and identically distributed random variables chosen from a probability distribution $P(s_{j}^{\pm } )$, given by (see inset to Fig.~\ref{fig1}),
\begin{align} \label{GrindEQ__5_}
{P(s)}  {=} \begin{cases}
\displaystyle \frac{f}{1-u} & \mbox{for $u<s<1$}, \\
&\\
\displaystyle \frac{1-f}{1-u} & \mbox{for $-1<s<-u$}, \\
&\\
0,& \mbox{otherwise}.
\end{cases}
\end{align}
The parameter $u$, $0<u\le 1$, controls the width of the positive and negative parts of the distribution, while $f$, $0<f<1$, determines the ratio of inhibitory to excitatory connections.  This functional form excludes connections that are very close to zero, which would bias the 1d network towards falling apart into disjoint pieces.   The coupling $g$ in Eq.~\eqref{GrindEQ__4_} controls the strength of a systematic clockwise ($g>0$) or counterclockwise ($g<0$) bias in the strengths of positive and negative neural connections around the ring.  As we shall see, nonzero $g$ can have a remarkable effect on the spectrum and localization properties. In this paper, we concentrate on the spectra and localization properties of eigenfunctions in the approximately balanced case, $f\approx 1/2$, which represents the greatest departure from conventional Hermitian localization problems in one dimension~\cite{anderson, efros,chaudhuri2014diversity, ziman, thouless}.    For now, we suppress the neuroscience constraints associated with Dale's Law, as might be appropriate if each node in the chain describes a large number of strongly coupled neurons randomly chosen to be excitatory and inhibitory.   However, we shall later argue (Sec.~\ref{dale_sect}) that a straightforward modification of Eq.~\eqref{GrindEQ__4_} that respects Dale's Law produces negligible changes in the spectra and localization properties in the limit of large $N$.

 The case of $f=1$ with $0<u<1$ (random excitatory connections only) is related to earlier work on the random non-Hermitian 1d matrices that arise from the physics of randomly pinned superconducting vortex lines~\cite{hatano1997vortex, goldsheid1998distribution} and in the population dynamics of heterogeneous 1d environments~\cite{shnerb1,shnerb2}. When $g>0$, this problem is sometimes referred to as ``directed localization''~\cite{efetov1997directed, brouwer1997}, terminology we adopt here as well.  The spectrum of models with $f=1/2$ and $u=1$ (\textit{i.e}., $\boldsymbol M(j,j+1)=\pm 1$ and  $\boldsymbol M(j+1,j)=\pm 1$, excitatory/inhibitory connections chosen at random with equal probabilities) has also been studied before~\cite{zee1, zee2, model1, model2, model3,hagger,hagg}, and has been shown to have an extremely rich structure.  Here, we explore the \textit{localization} properties of the eigenfunctions associated with these spectra for a range of $u$ values in some detail.

\begin{figure}[t]
\includegraphics[width=.43\textwidth]{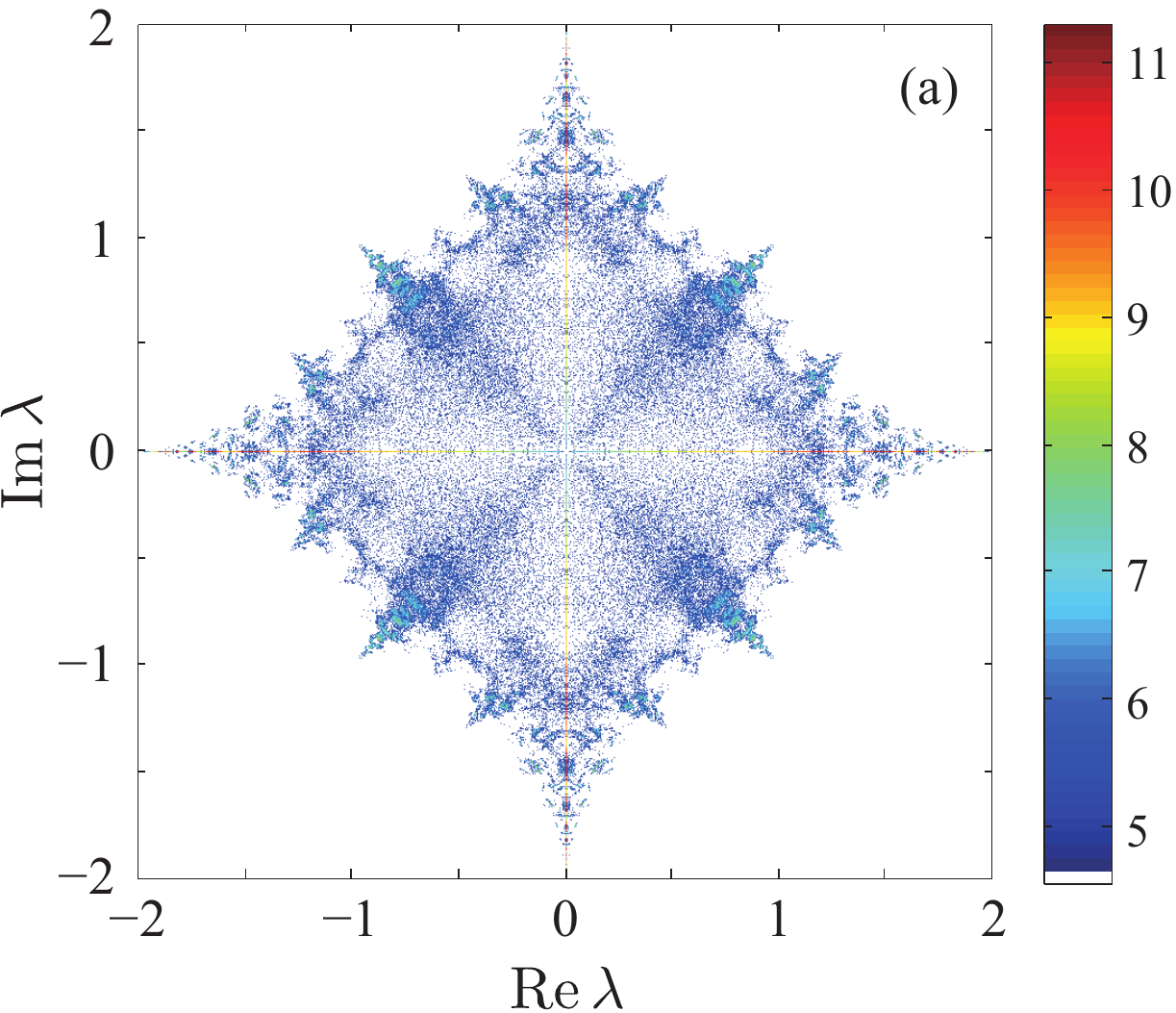}

\vspace{\baselineskip}
\includegraphics[width=.43\textwidth]{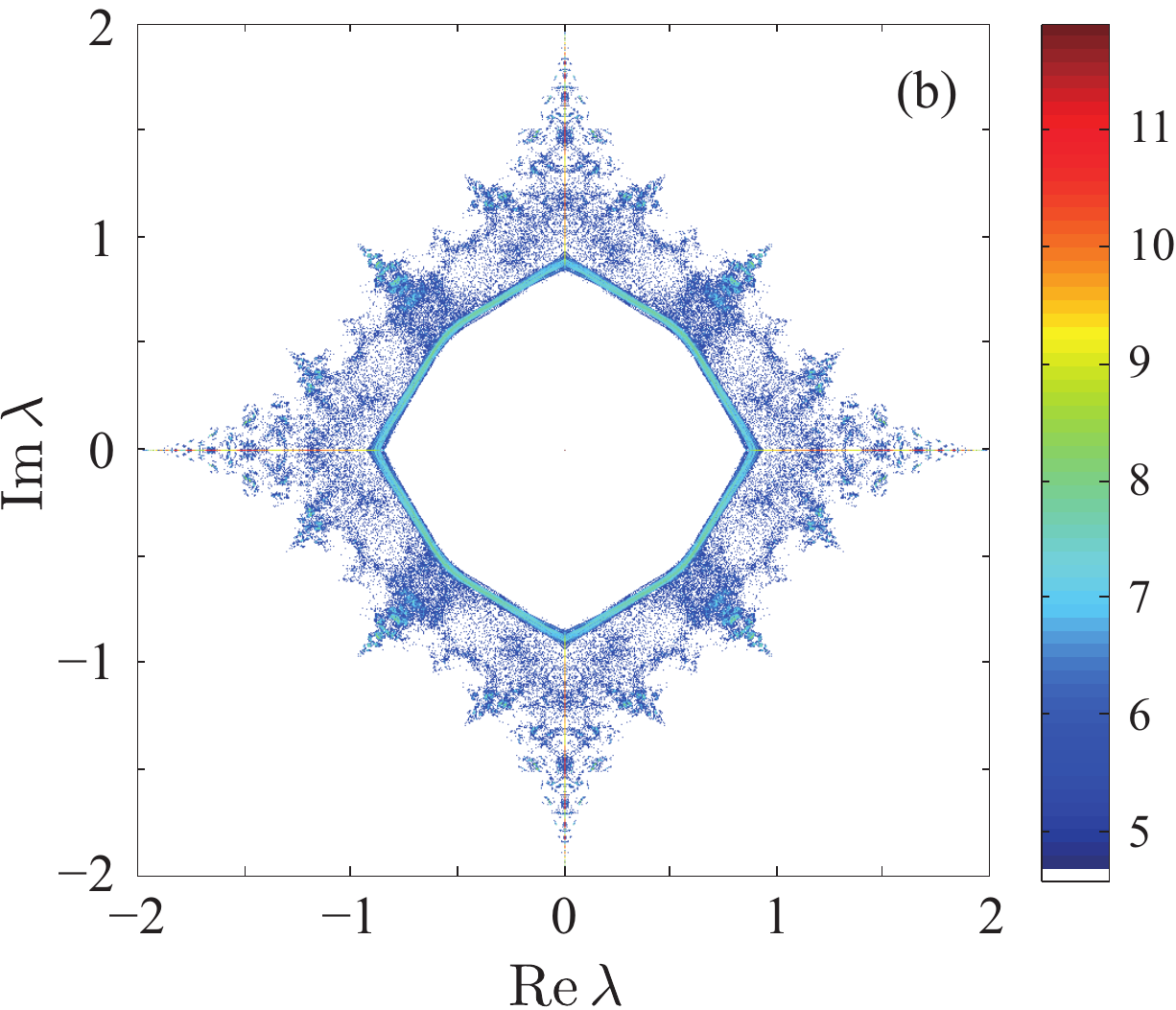}
\caption {(a) Density-of-states (DOS) of the complex eigenvalue spectrum for the sparse random matrix defined by $f = \frac{1}{2}$ and $u = 1$ with $g = 0$  obtained through exact diagonalization of 10,000 matrices of dimension $5000 \times 5000$. The scale is logarithmic, and the white background denotes areas where the DOS vanishing. The only randomness is in the signs of the connections, $\boldsymbol M(j,j+1) = \pm 1$, $\boldsymbol M(j+1,j) = \pm 1$.
  (b) Density-of-states for the case $g=0.1$, all other parameters are identical to (a).   }
\label{dos}
\end{figure}


\begin{figure}
\includegraphics[width=.45\textwidth]{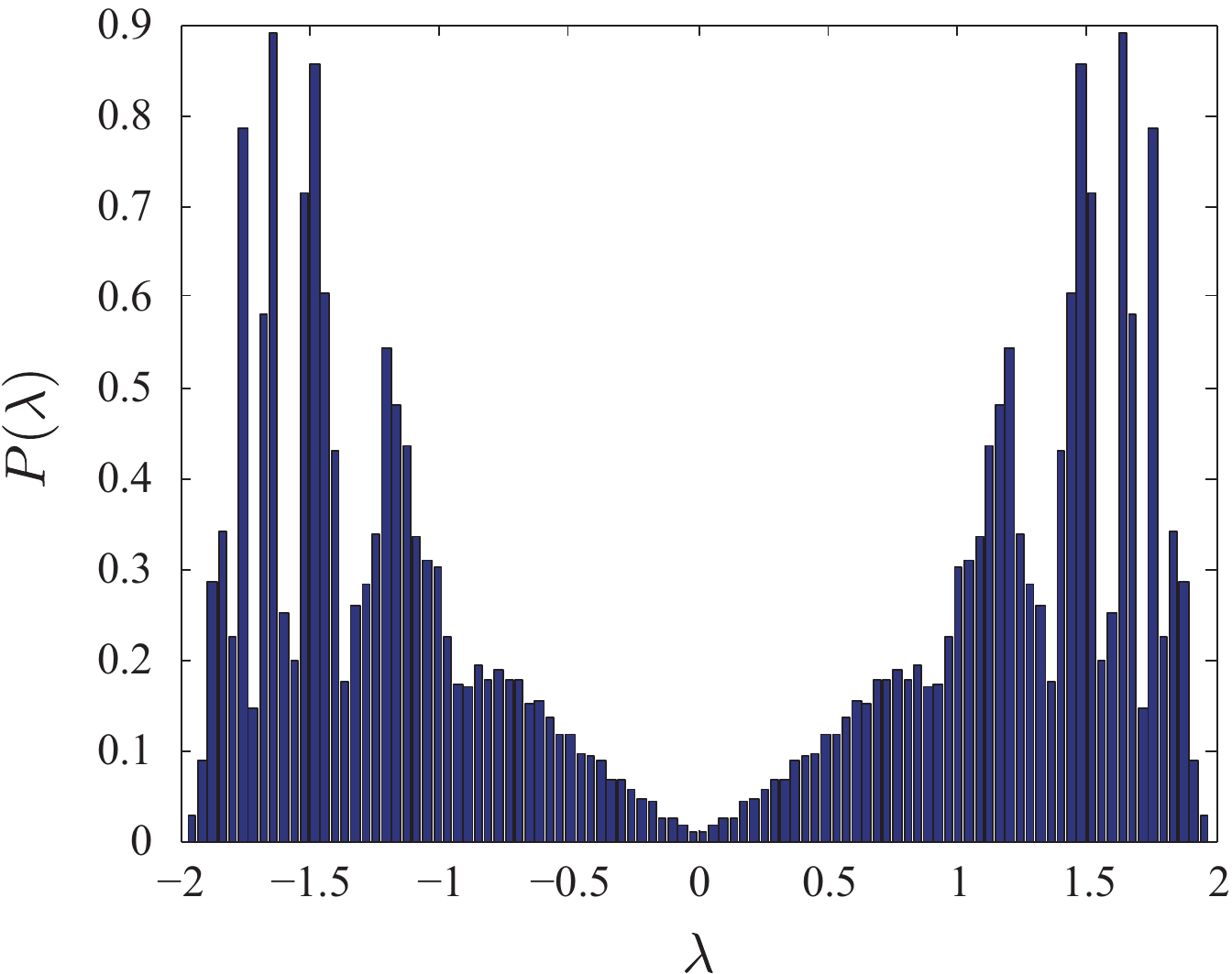}
\caption {DOS on the real axis for the parameters of Fig.~\ref{dos}(a), comprising close to 20\% of all eigenvalues.}
\label{dos2}
\end{figure}

Figures~\ref{dos}(a) and~(b) exhibit the remarkable spectra associated with Eqs.~\eqref{GrindEQ__4_} and~\eqref{GrindEQ__5_} when   $f=1/2$, $u=1$ and for $g=0$ and $g=0.1$ respectively. To the best of our knowledge, the striking spectrum in Fig.~\ref{dos}(a) first appeared in a 1999 paper by Feinberg and Zee~\cite{zee1}, who mentioned that this model might have interesting localization properties. Although the eigenvalues are in general complex, when $g=0$ a significant fraction of them (about 20\%) have condensed onto the real axis, see Fig.~\ref{dos2}; similarly, about 20\% have condensed onto the imaginary axis. In Appendix~\ref{app-h2} this numerical analysis is extended to the case of $u \neq 1$, with similar results. For large $N$, the density-of-states is symmetric under reflections across the real and imaginary axes, as well as across $\pm 45^\circ$  lines  in the complex plane, as we shall show in Sec.~\ref{symmetries}. The remaining eigenvalues (approximately 60\%) form an intricate, diamond-shaped structure.  When $u$ is near $1$, the density-of-states appears to acquire a fractal-like boundary. See Appendix B for a summary of the density-of-states for the more general probability distribution of Eq.~\eqref{GrindEQ__5_} for arbitrary $u$ and $f=1/2$.

\subsection{Main Results}

We are now in a position to summarize our main results. In Sec.~\ref{symmetries} we discuss various symmetries associated with the density-of-states of the models studied here. Sec.~\ref{loc_sect} we show that almost all eigenfunctions are localized (similar to the 1d Hermitian case of Appendix A), with the smallest localization lengths near the boundary of the spectrum in the complex plane, and a diverging localization length near the origin.  Our analysis of localization in this model has been guided by work of Derrida \textit{et al}.~\cite{derrida} on a related problem (with unimodular \textit{complex} random couplings between sites), who derive an elegant generalization of the Thouless formula for eigenvalues in the complex plane:  The inverse localization length is the two-dimensional electrostatic potential associated with a collection of charges at the eigenvalue locations in the complex plane.  Our numerical analysis strongly suggests that the localization length diverges as the modulus of the eigenvalues tends to zero.   Indeed, if the eigenvalues are written $\lambda =\lambda _{1} +i\lambda _{2} $, where $\lambda _{1}$ and $\lambda _{2}$ are real, a numerical study of the inverse localization length defined via the product of $N$ random $2 \times 2$ transfer matrices~\cite{furstenberg1963noncommuting, ishii, crisanti1993products} for $u$ near $1$ leads to the following ansatz:
\begin{align} \label{GrindEQ__6_}
\xi (\lambda _{1} ,\lambda _{2} )\propto \frac{1}{(\left|\lambda _{1} \right|+\left|\lambda _{2} \right|)\sqrt{\lambda _{1}^{2} +\lambda _{2}^{2} } } ,
\end{align}
which should be compared to the much weaker logarithmic divergence discussed in Appendix A for 1d Hermitian hopping randomness. Although the divergence shown in Eq.~\eqref{GrindEQ__6_} only holds for $u$ near 1, the infinite localization length at the origin is more general, as discussed in Sec.~\ref{vanish}.

%


As shown in Fig.~\ref{dos}(b), a hole surrounding the origin with angular corners opens up in the complex plane when these calculations are repeated for a clockwise bias parameter $g=0.1$. A similar hole opens up in the Feinberg-Zee model of Ref.~[\onlinecite{zee1}], with complex unimodular hopping matrix elements~\cite{molinari}.    As we demonstrate in Sec.~\ref{loc_sect}, a large number of \textit{extended} states now occupies the rim of hole.  For large $N$, the eigenvalues of the localized states outside the hole are unchanged, a spectral rigidity property that can be derived from a simple exponential ``gauge transformation'' acting on the corresponding eigenfunctions for $g=0$, see Ref.~[\onlinecite{hatano1997vortex}]. A corollary is that the $g$-dependent shape of holes like that in Fig.~\ref{dos}(b) tracks the $g = 0$ contours of constant localization length, with a diverging localization length as the rim is approached from the outside.

%

 What happens to directed localization for neural networks that obey Dale's Law, as studied for spatially extended neural connections in Ref.~[\onlinecite{abbott}]?   In Sec.~\ref{dale_sect}, we argue that the above results should be unchanged for large $N$. We will do this by replacing the matrix $\boldsymbol M$ with a modified connectivity matrix,
\begin{align} \label{GrindEQ__9_}
&\boldsymbol G=\sum _{k=1}^{N}\sigma _{k} \left[e^{g} {\left| k+1 \right\rangle} {\left\langle k \right|} {\rm \; +\; }e^{-g} {\left| k-1 \right\rangle} {\left\langle k \right|} \right] ,\nonumber\\
&{\left| k+N \right\rangle} ={\left| k \right\rangle}
\end{align}
where the $N$ real random numbers $\{ \sigma _{k} \} $ are chosen from the probability distribution $P(\sigma _{k} )$, again given by Eq.~\eqref{GrindEQ__5_}.
Figure~\ref{dale} illustrates a particular example of the Dale's Law connectivity matrix for  $N=5$.    Note that the nonzero connections in the same row have the same sign.  Equation~\eqref{GrindEQ__9_} has site randomness, as opposed to the bond randomness displayed in Eq.~\eqref{GrindEQ__4_}.   Despite the fact that $2N$ random numbers are necessary to specify $\boldsymbol M$ and only $N$ random numbers specify $\boldsymbol G$, we show via similarity transformations in Sec.~\ref{dale_sect} that the spectra and localization properties of $\boldsymbol M$ and $\boldsymbol G$ are essentially identical, a result which we have also confirmed numerically.   The underlying reason is that the spectral properties are determined in both cases by above/below diagonal products such as $\boldsymbol M(j,j+1)\cdot \boldsymbol M(j+1,j)=s_{j}^{+} s_{j}^{-} $ and  $\boldsymbol G(j,j+1)\cdot \boldsymbol G(j+1,j)=\sigma _{j} \sigma _{j+1} $.    These quantities have identical statistical properties.

In Sec.~\ref{perturb_sect} we discuss large-$g$ perturbation theory, that focuses on the changes in the eigenvalues and eigenfunctions which are all extended in this limit. This analysis can be carried out for arbitrary $f$ and $u$, although it cannot capture the localization that results as the eigenvalues move toward the origin with decreasing $g$.

Section~\ref{sum_sect} contains a summary and outlook, including a brief discussion of the effect of diagonal randomness. Appendices A-C describe, respectively, a Hermitian random hopping model, the density-of-states for arbitrary $u$, and second-order perturbation theory for large $g$.

\section {Symmetries associated with the density-of-states}
\label{symmetries}

In order to discuss spectral symmetries, we first introduce a similarity transformation which is applicable to the present model with  open boundary conditions.
Consider an $N\times N$ tridiagonal matrix of the form
\begin{align}\label{eq-h10}
\boldsymbol A=\begin{pmatrix}
d_1 & b_1 &    &  &  & 0 \\
a_1 & d_2 & b_2 \\
& a_2 & d_3 & b_3 && \\
&& \ddots & \ddots & \ddots  \\
&& & a_{N-2} & d_{N-1} & b_{N-1} \\
0 & &  & &  a_{N-1} & d_N
\end{pmatrix},
\end{align}
where $\{a_x\}$, $\{b_x\}$ and $\{d_x\}$ are arbitrary real numbers, and all other entries vanish.
We can symmetrize this matrix by a diagonal similarity transformation, whose $j$th matrix element reads
\begin{align}\label{eq-h20}
\boldsymbol S_{jj}=\prod_{k=1}^{j-1} \sqrt{\frac{a_k}{b_k}},
\end{align}
which we may call a generalized gauge transformation~\cite{hatano1997vortex}.
The result of this symmetrization reads
\begin{align}\label{eq-h30}
\boldsymbol A' = \boldsymbol S^{-1}\boldsymbol A \boldsymbol S=\begin{pmatrix}
d_1 & c_1 &&& & 0 \\
c_1 & d_2 & c_2 && \\
& c_2 & d_3 & c_3 & \\
& &\ddots & \ddots & \ddots & \\
&& & c_{N-2} & d_{N-1} & c_{N-1} \\
0&&& &  c_{N-1} & d_N
\end{pmatrix},
\end{align}
where
\begin{align}\label{eq-h40}
c_j=\sqrt{a_j b_j}.
\end{align}
The matrices~\eqref{eq-h10} and~\eqref{eq-h30} are isospectral, due to the properties of similarity transformations.

Note that the spectrum of the matrix $\boldsymbol A'$ depends only on the product of the opposing off-diagonal elements $a_j$ and $b_j$, not independently on each of them (as also follows from calculating the characteristic polynomial of the matrix $\boldsymbol A$).
Another important observation is that the matrix~\eqref{eq-h30} is real and symmetric if $a_j$ and $b_j$ have the same sign, but is non-Hermitian otherwise.
The non-Hermitian Anderson chains proposed in Refs.~[\onlinecite{hatano1996localization,hatano1997vortex}], in which all $a_j$ and $b_j$ are negative, therefore would have only real eigenvalues unless we introduced periodic boundary conditions.
If the matrix~\eqref{eq-h10} has non-zero corner elements $a_N$ for $\boldsymbol A_{1N}$ and $b_N$ for $\boldsymbol A_{N1}$ (thus coupling the chain into a ring), the resulting matrix $\boldsymbol A'= \boldsymbol S^{-1}\boldsymbol A\boldsymbol S$ has non-zero corner matrix elements
\begin{align}\label{eq-h50}
\boldsymbol A'_{1N}=\left(\boldsymbol S^{-1}\boldsymbol A\boldsymbol S\right)_{1N}&=\sqrt{a_Nb_N}\prod_{j=1}^{N}\sqrt{\frac{a_j}{b_j}},
\\\label{eq-h60}
\boldsymbol A'_{N 1}=\left(\boldsymbol S^{-1}\boldsymbol A\boldsymbol S\right)_{N1}&=\sqrt{a_Nb_N}\prod_{j=1}^{N}\sqrt{\frac{b_j}{a_j}},
\end{align}
which make the matrix non-Hermitian and allows the possibility of complex eigenvalues unless
\begin{align}\label{eq-h70}
\prod_{j=1}^{N}a_j=\prod_{j=1}^Nb_j.
\end{align}

Although this similarity transformation leaves the diagonal randomness intact, it packs all effects of the random, non-Hermitian hopping terms into a single pair of corner matrix elements.   This perspective is useful already for simple cases where the elements $\{ a_{j} \}$ and $\{ b_{j} \}$ in Eq.~\eqref{eq-h10} can be  different, but are both constrained to be of the same sign.  Let us take $a_{j} =e^{-g} s_{j}^{-} >0$ and $b_{j} =e^{g} s_{j}^{+} >0$, consistent with Eq.~\eqref{GrindEQ__4_}, so that the corner matrix element $\boldsymbol A'_{N1}$ takes the form
\begin{align}
 \boldsymbol A'_{N1} =\sqrt{s_{N}^{-} s_{N}^{+} } e^{Ng} \prod _{j=1}^{N}\sqrt{\frac{s_{j}^{+} }{s_{j}^{-} } }. \label{eqA}
\end{align}
If we now choose the elements $\left\{s_{j}^{\pm } \right\}$ according to the probability distribution of Eq.~\eqref{GrindEQ__5_} with $f=1$ and $0<u<1$, all $s_{j}^{\pm } $ will be positive, and $\boldsymbol A'_{N1}$ is real and described by a log-normal distribution.    It is simpler to study $\log \boldsymbol A'_{N1} $, which behaves like a random walk.  As discussed in Sec.~\ref{loc_sect}, a closely related quantity determines the localization properties of eigenfunctions as function of $g$.  Focusing for simplicity on the case $u=0^{+} $, we readily find
\begin{align}
\langle \log \boldsymbol A'_{N1} \rangle =Ng-1,
\end{align}
and
\begin{align}
 \left\langle {\rm \; }\left(\log \boldsymbol A'_{N1} -\left\langle \log \boldsymbol A'_{N1} \right\rangle \right)^{2} \right\rangle =\frac{N}{4} +O(1/N),
\end{align}
where $\left\langle \bullet \right\rangle $ represents an average over the disorder and similar results obtain for $0<u<1$.  Upon defining an effective directional bias parameter $g_\mathrm{eff} \equiv \left\langle \log \boldsymbol A'_{N1} \right\rangle /N$, we see that if microscopic bias is $g=0$, then the hopping randomness represented by the elements $\left\{s_{j}^{\pm } \right\}$ leads to a $g_\mathrm{eff} =O(1/N^{1/2} )$, which vanishes in the limit large $N$.  Thus, the hopping disorder is effectively undirected as $N\to \infty $ in this case.  When diagonal randomness is also present, we expect that the localized states will remain localized with real eigenvalues, unless $g$ exceeds a critical value $g_{c1} $ given by the minimum inverse localization length when $g=0$.

If $a_j$ and $b_j$ can have different signs, the matrix~\eqref{eq-h10} is inherently non-Hermitian and can have complex eigenvalues with or without periodic boundary terms.
It is this interesting case we focus on in the present paper. As discussed below, coupling the chain into a ring is crucial when $g>0$.

\subsection{Spectrum of sign-random model}
\label{subsec-h2}

\begin{figure}
\includegraphics[width=0.23\textwidth]{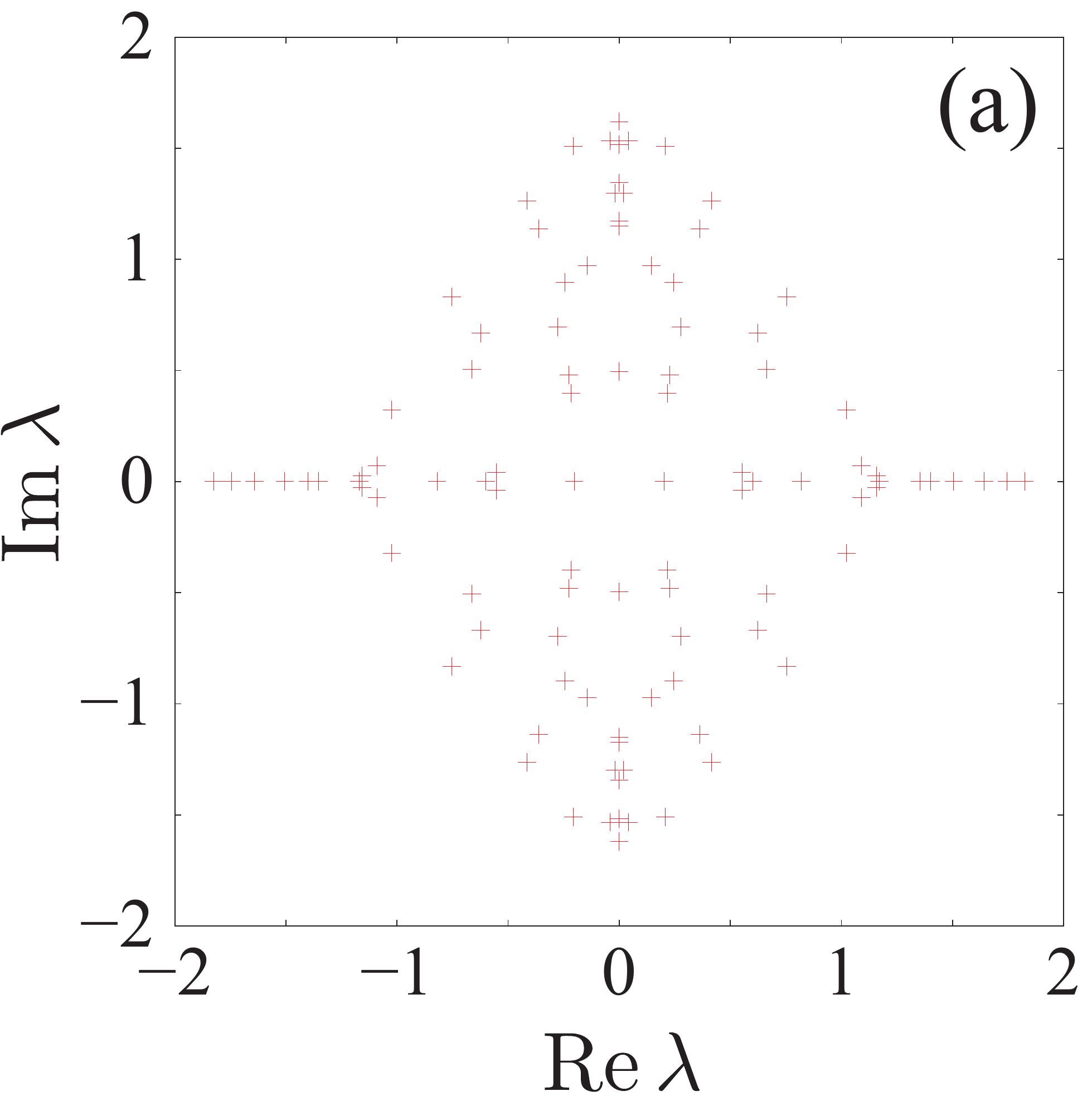}

\vspace*{\baselineskip}
\includegraphics[width=0.23\textwidth]{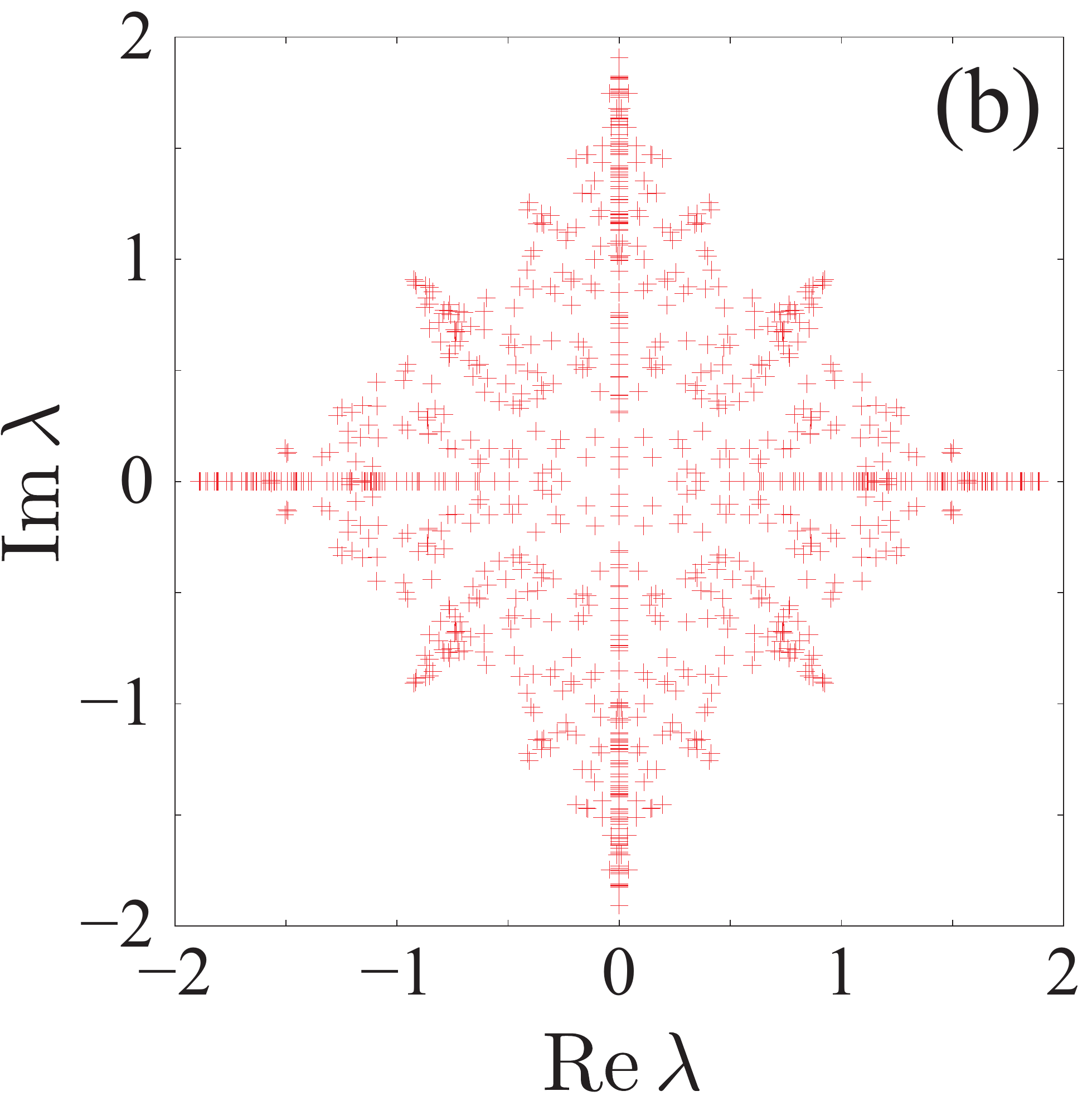}

\vspace*{\baselineskip}
\includegraphics[width=0.23\textwidth]{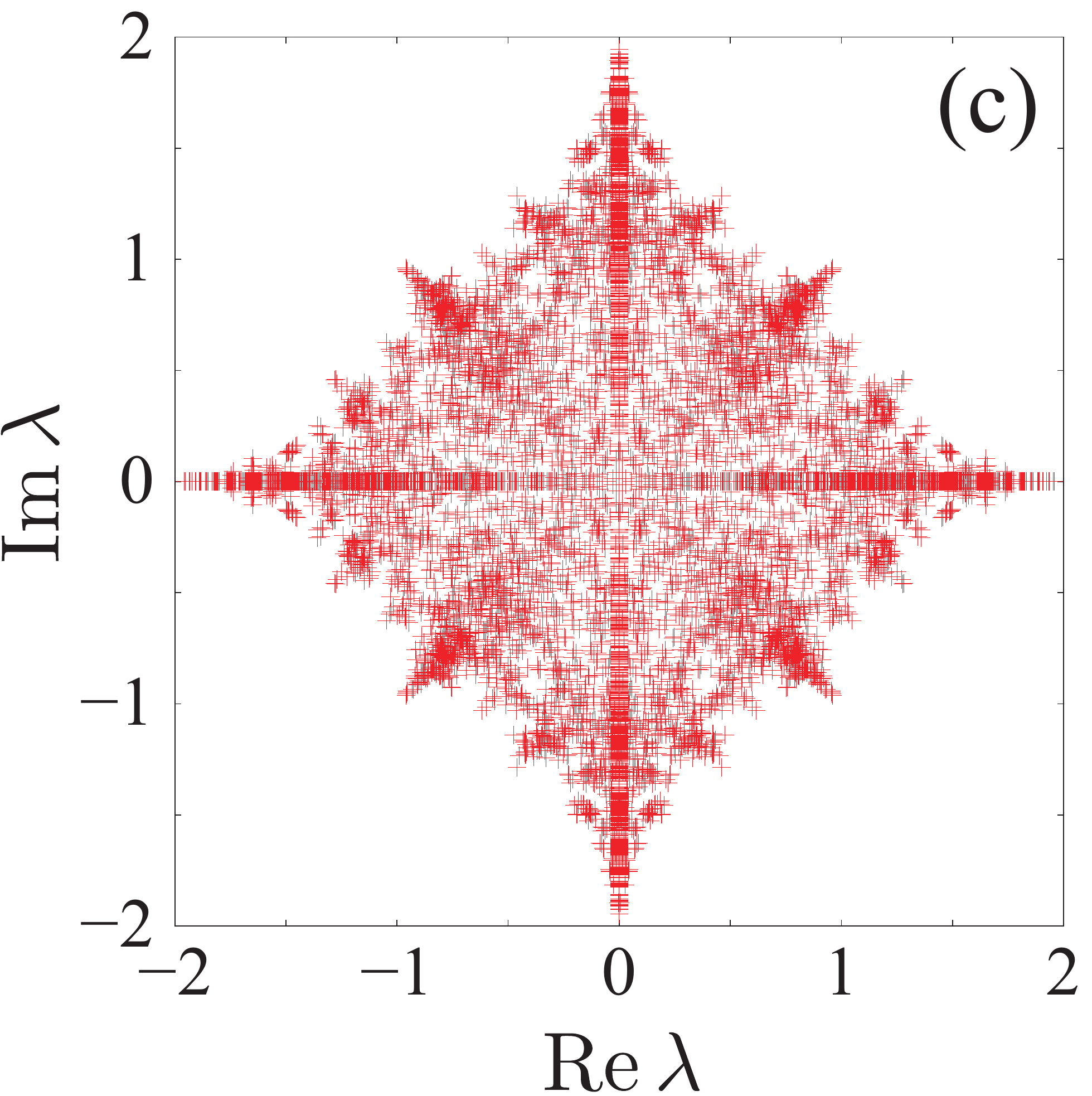}
\caption{The spectra on complex planes of the matrix~\eqref{eq-h80} for (a) $N=100$, (b) $N=1000$ and (c) $N=10,000$.}
\label{fig-h1}
\end{figure}

Let us apply the above considerations to the sign-random non-Hermitian tight-binding chain given by the $N\times N$ matrix corresponding to Eq.~\eqref{GrindEQ__4_} with $g=0$:
\begin{align}\label{eq-h80}
\boldsymbol M=
\begin{pmatrix}
 & s^+_1 &  &  &  &  &  \\
s^-_1 &  & s^+_2 &   &  &  &  \\
 & s^-_2 &  & s^+_3 &  &  &  \\
 &  & s^-_3 &  & \ddots & &  \\
 &  &   &\ddots  &  & \ddots & \\
 &  &  &&\ddots && s^+_{N-1} \\
 &  &  &  & & s^-_{N-1} & \\
\end{pmatrix},
\end{align}
where all remaining elements, including diagonal ones, vanish, and each of $s^\pm_j$ are randomly set to be $\pm 1$ with probability $1/2$.
Spectra found by numerical diagonalization of a random sample are shown in Fig.~\ref{fig-h1} for $N=10$, $1000$ and $10,000$, which should be compared with the disorder-averaged spectrum shown in Fig.~\ref{dos}(b). As discussed below, when $g=0$ we can neglect the corner matrix elements.

We can describe the symmetries of the spectrum in the following way:
First, since the matrix~\eqref{eq-h80} is real, namely, $\boldsymbol M=\boldsymbol M^\ast$, if there is an eigenvalue $\lambda_n$, there must be another eigenvalue $\lambda_n^\ast.$
In other words, the spectrum is symmetric with respect to reflections across the real axis.
Second, since the spectrum depends only on the product of $s_x^+$ and $s_x^-$, the matrices $H$ and $-H$ are isospectral, and hence if there is an eigenvalue $\lambda_n$, there must be another eigenvalue $-\lambda_n$.
In other words, the spectrum is symmetric under inversion in the complex plane $\lambda \to -\lambda$. Upon combining the two symmetries, we see that the matrices $\boldsymbol M$ and $-\boldsymbol M^\ast$ are isospectral, and hence the spectrum is symmetric with respect to reflections around the imaginary axis, too. This argument holds in a statistical sense even if we add zero-mean diagonal randomness into Eq.~\eqref{eq-h80}.

Finally, we argue that the spectrum has statistical symmetry with respect to the reflections across the $45^\circ$ lines $\mathop{\mathrm{Re}}E=\pm\mathop{\mathrm{Im}}E$.
According to the argument in the beginning of this section, the spectrum depends only on whether the product $s_x^+s_x^-$ is $+1$ or $-1$.
In other words, the randomness of the matrix~\eqref{eq-h80} is caused by independent probability distributions of $N-1$ independent degrees of freedom, $\{s_x^+s_x^-\}$, instead of $2N-1$.
Let us then consider the spectrum of the matrix $i\boldsymbol M$. By multiplying every matrix element by $i=\sqrt{-1}$, we flip the sign of the product of the opposing off-diagonal elements which, however, does not change the binomial distribution of the $N-1$ pieces of random variables when $f=1/2$.
Therefore, the matrices $\boldsymbol M$ and $i\boldsymbol M$ are statistically isospectral.
Since the spectrum of $i\boldsymbol M$ is given by the $90^\circ$ rotation of that of $\boldsymbol M$, the spectrum is statistically symmetric with respect to this operation.
Combining this symmetry with the other symmetries, we conclude that it is statistically symmetric with respect to reflections around the $45^\circ$ lines.
The fact that the symmetry becomes better as we increase the system size underlines the observation that the symmetry is indeed statistical.

\begin{figure}
\includegraphics[width=0.23\textwidth]{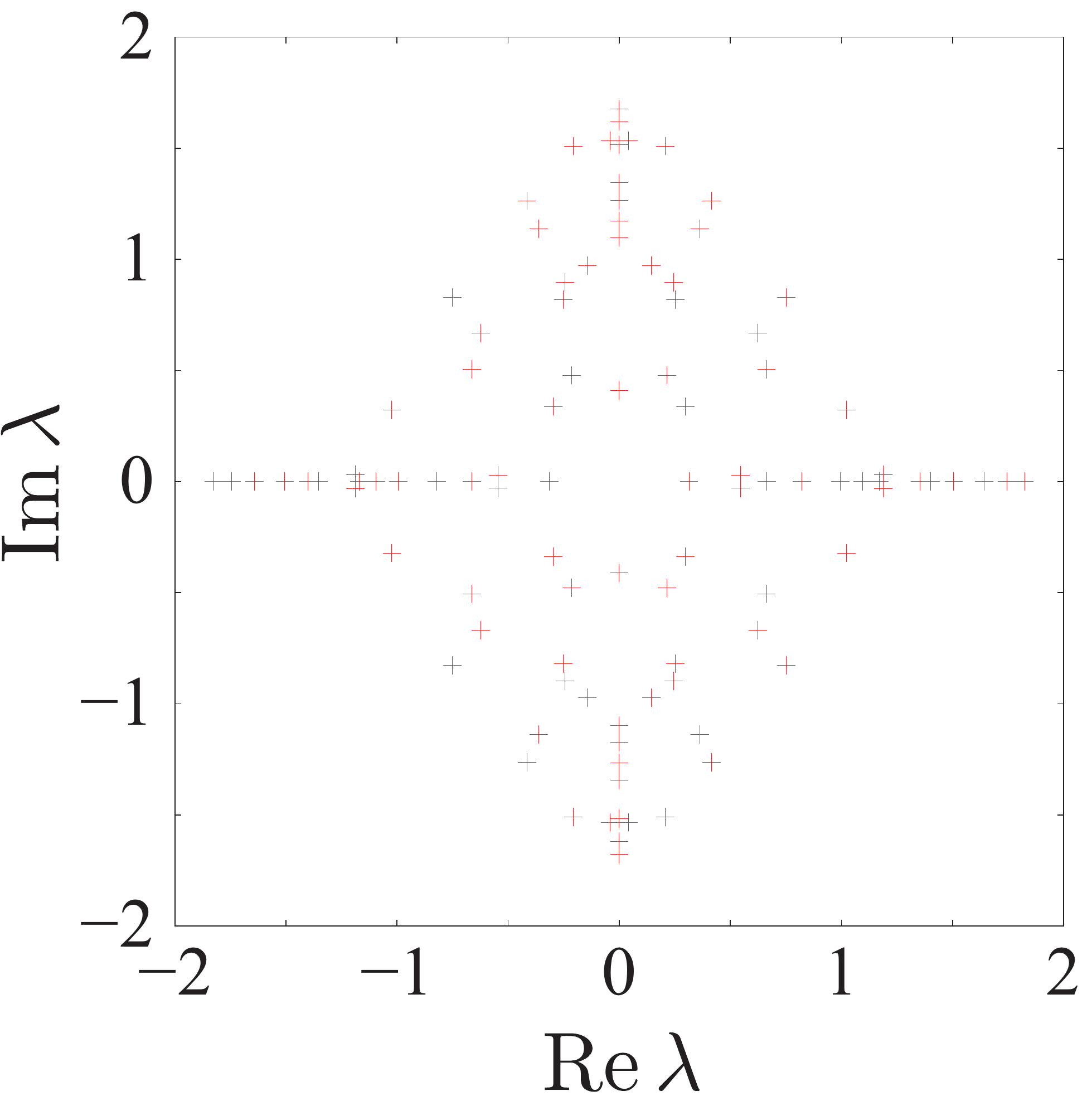}
\caption{The spectrum on a complex plane of the matrix~\eqref{eq-h80}, this time \textit{with} additional boundary elements; compare it with Fig.~\ref{fig-h1}(a). The system size is $N=100$.}\label{fig-h2}
\end{figure}

Adding the boundary elements $\boldsymbol M_{1N}=s^+_N$ and $\boldsymbol M_{N1}=s^-_N$ does not change the spectrum in an essential way when $g=0$;
their first-order perturbation to an eigenvalue $\lambda_n$ with its normalized left- and right-eigenvectors $\langle \tilde{\psi}_n|$ and $|\psi_n\rangle$ (we use the tilde symbol to emphasize that they are not Hermitian conjugate to each other) is of order $1/N$ at most (and is exponentially small if the eigenfunctions are localized).
Indeed, comparison of numerical results of Figs.~\ref{fig-h1}(a) and~\ref{fig-h2} with and without the boundary elements suggests that they are not only statistically the same but also almost identical with occasional differences, even for $N=100$.

\subsection{Asymmetric amplitudes}
\label{subsec-h3}

Let us next introduce asymmetric amplitudes to the sign-random tight-binding model.
Following Refs.~[\onlinecite{hatano1996localization,shnerb1,hatano1997vortex}], we express the asymmetry in the form (equivalent to Eq.~\eqref{GrindEQ__4_})
\begin{align}\label{eq-h90}
\boldsymbol M(g)=
\begin{pmatrix}
 & e^{g}s^+_1 &   &  &  & e^{-g}s^-_N \\
e^{-g}s^-_1 &  & e^{g}s^+_2 &   &  &  \\
 & e^{-g}s^-_2 &  &  &  &  \\
 &   \ddots &  & \ddots & &  \\
 &     &\ddots  & &  \ddots & \\
 &    &&\ddots & & e^{g}s^+_{N-1} \\
e^{g}s^+_N &  &    & & e^{-g}s^-_{N-1} & \\
\end{pmatrix},
\end{align}
where we assume $g>0$ without loss of generality.
Note here that we have included the boundary terms $e^{g}s^+_N$ and $e^{-g}s^-_N$;
if not, the spectrum would be $g$-independent because it would depend only on the product of the opposing off-diagonal elements.
The diagonal similarity transformation
\begin{align}\label{eq-h100}
\boldsymbol T(g)_{xx}=e^{-g(x-1)}
\end{align}
changes the matrix $\boldsymbol M(g)$ into
\begin{align}\label{eq-h110}
&\boldsymbol M'(g) = \boldsymbol T(g)^{-1}\boldsymbol H(g)\boldsymbol T(g)= \nonumber \\
& \hspace{2 cm} \begin{pmatrix}
 & s^+_1 &  & & & e^{-Ng}s^-_N \\
s^-_1 &  & s^+_2 &  &  \\
 & s^-_2 &  & &  \\
 &  & \ddots& & \ddots   \\
  &  & & && s^+_{N-1} \\
e^{Ng} s^+_N & & & &    s^-_{N-1} & \\
\end{pmatrix},
\end{align}
which shows that the boundary elements are essential in having a strong dependence on $g$.

\begin{figure}
\includegraphics[width=0.23\textwidth]{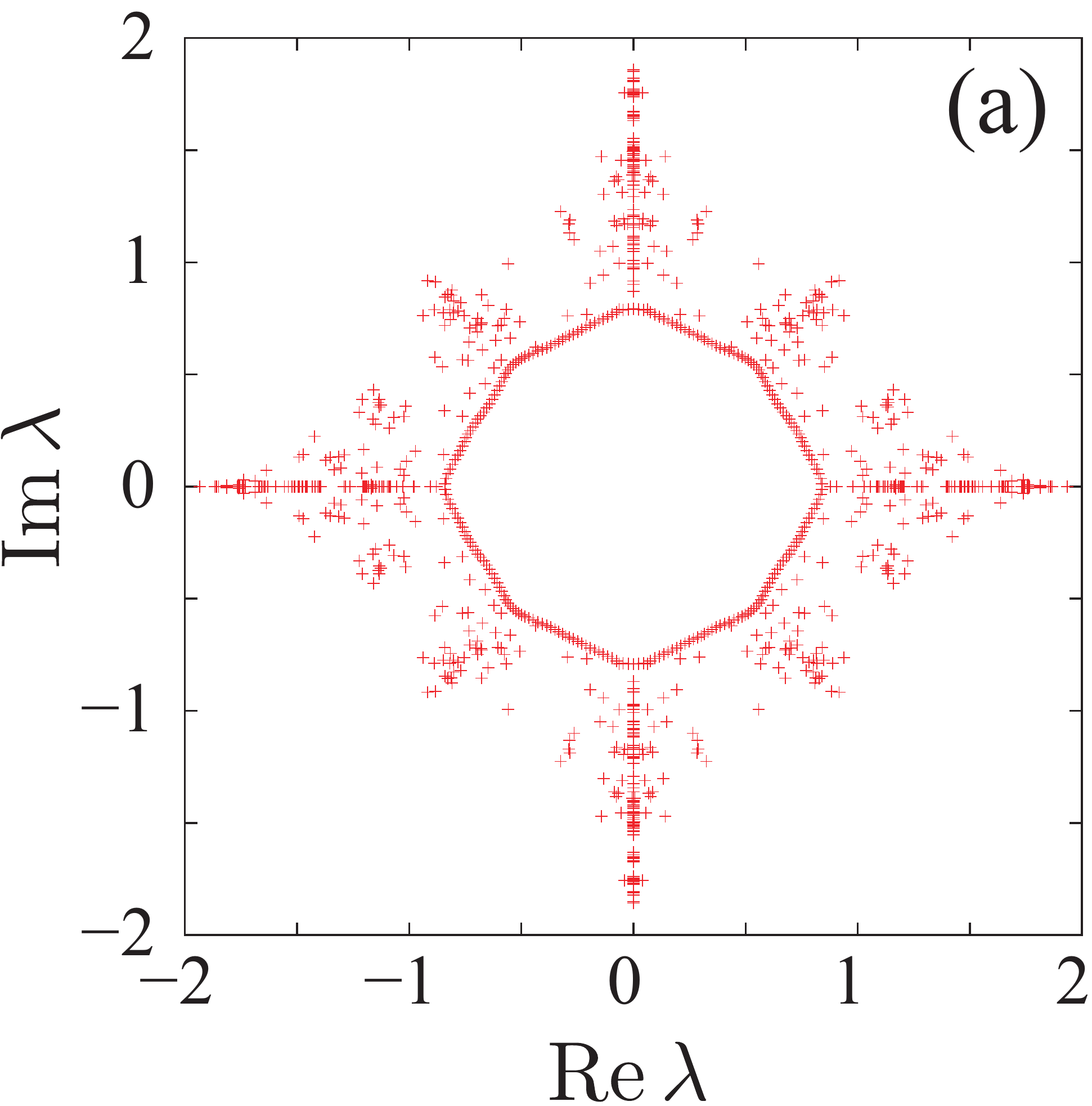}

\vspace{\baselineskip}
\includegraphics[width=0.23\textwidth]{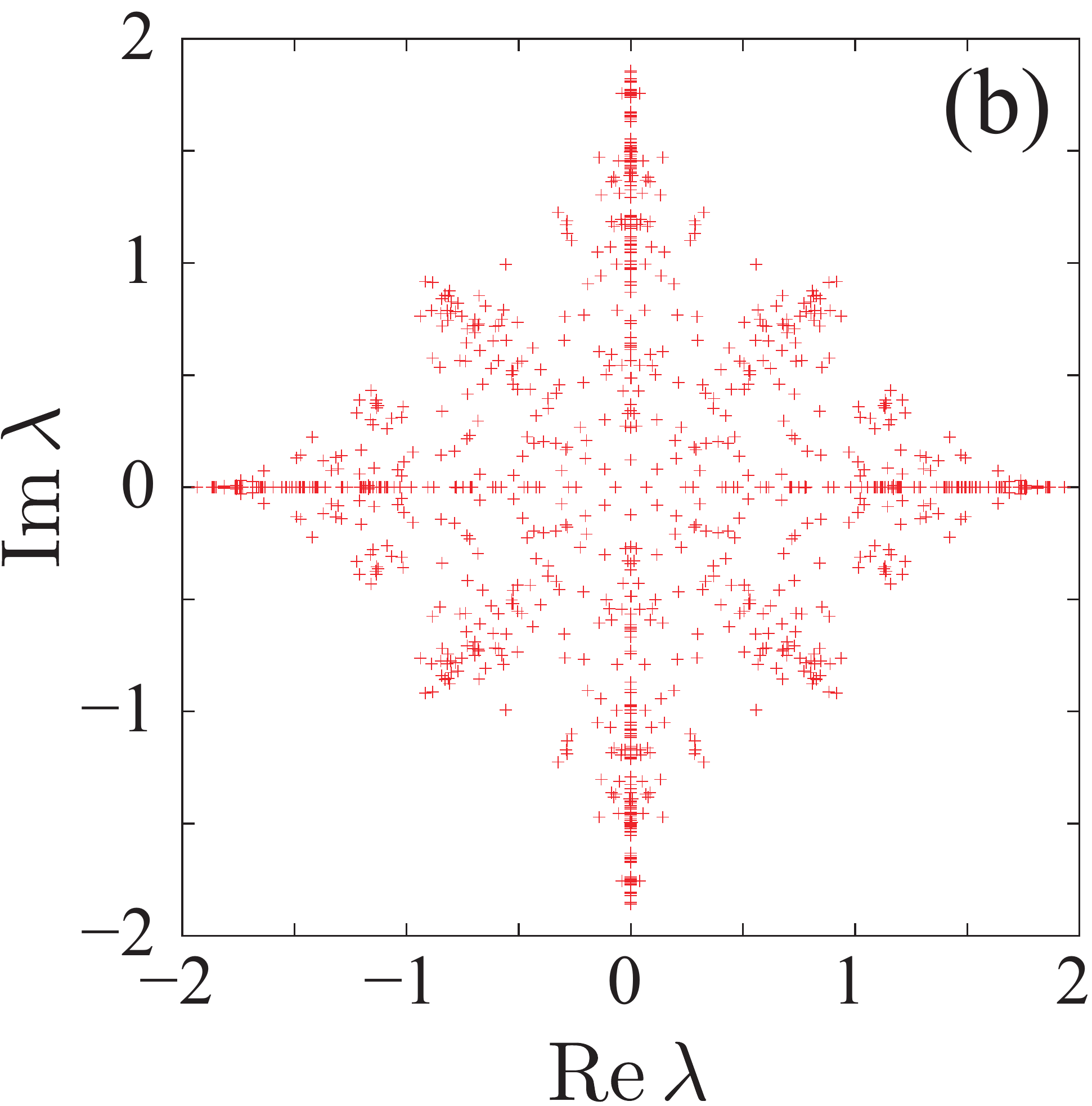}

\vspace{\baselineskip}
\includegraphics[width=0.23\textwidth]{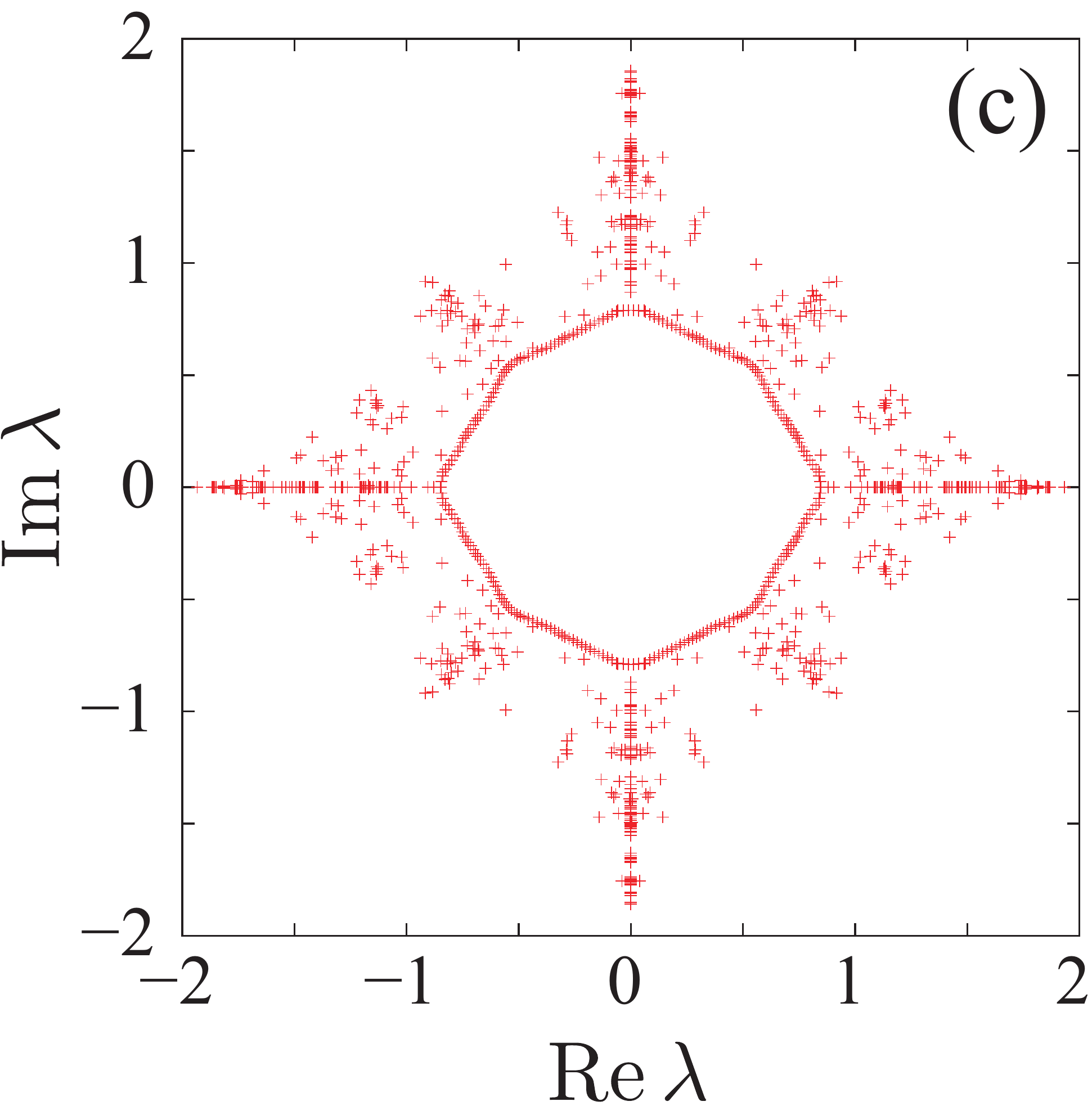}
\caption{The spectra on complex planes of (a) the directed matrix~\eqref{eq-h90} with random signs for $g=0.1$ and $N=1000$;
(b) the matrix~\eqref{eq-h140} obeying Dale's law for $N=1000$; Note the similarity to the spectrum in Fig.~\ref{fig-h1}(b), that does not have this restriction.
(c) the matrix~\eqref{eq-h150} for $g=0.1$ and $N=1000$. Note the close similarity with Fig.~\ref{fig-h3}(a).}
\label{fig-h3}
\end{figure}

As was discussed in Refs.~[\onlinecite{hatano1996localization,hatano1997vortex}], the spectrum of $\boldsymbol M(g)$ can in fact be an indicator of the localization of the eigenfunctions.
Suppose that the eigenfunction $\psi_n$ of an eigenvalue $\lambda_n$ of the original Hamiltonian $\boldsymbol M=\boldsymbol M(0)$ is localized around a site $x_0$ and behaves approximately as
\begin{align}\label{eq-h120}
\psi_n(x)\sim e^{-\kappa_n|x-x_0|}
\end{align}
except for a phase factor.
This quantity is also an approximate eigenfunction of $\boldsymbol T(g)^{-1}\boldsymbol M(g)\boldsymbol T(g)$, because the first-order perturbative corrections due to the boundary elements are exponentially small, of order $e^{-N(\kappa_n-g)}$,  when $g<\kappa_n$.
Thus, the corresponding eigenfunction of $\boldsymbol M(g)$ is given by
\begin{align}\label{eq-h130}
\psi_n(x;g)\sim e^{-gx-\kappa_n|x-x_0|}
\end{align}
except for a phase factor.
Indeed, the periodic boundary conditions are almost precisely satisfied for large $N$ if $g<\kappa_n$;
the discrepancy at the boundary is exponentially small, of order $e^{-N(\kappa_n-g)}$.
Therefore, the eigenvalue $\lambda_n$ of $\boldsymbol M(0)$ remains to be an eigenvalue of $\boldsymbol M(g)$ when $g<\kappa_n$.
This argument breaks down when $g>\kappa_n$, for which the eigenvalue now moves as a function of $g$, with motion starting when $g=\kappa_n$.
The numerical diagonalization of a random sample with $g=0.1$ gives Fig.~\ref{fig-h3}(a).
According to the above argument (elaborated in Sec.~\ref{loc_sect} in detail), the states on the inner curve similar to an octagon have $\kappa=0.1$ for $g=0$, and vanishing $\kappa$ for $g=0.1$.

\subsection{Spectrum of models obeying Dale's law}
\label{dale_sect}

\begin{figure}
\includegraphics[width=0.45\textwidth]{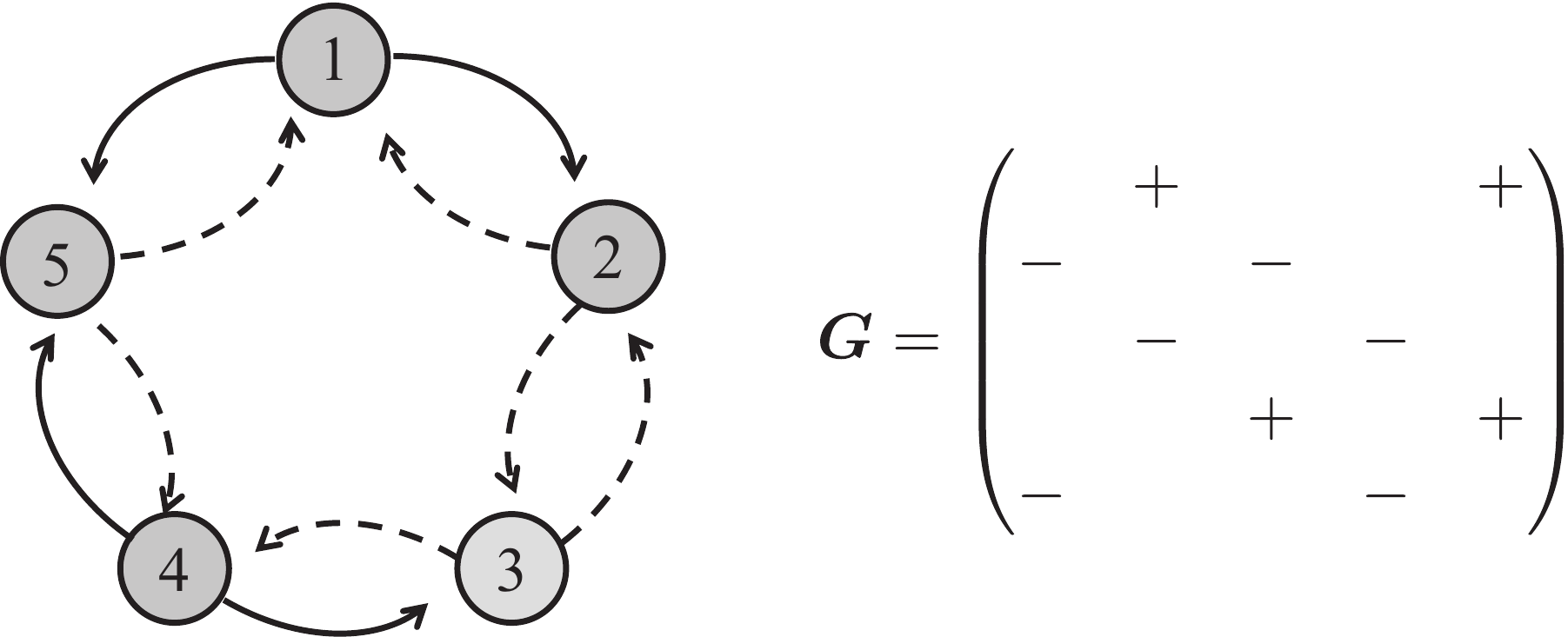}
\caption{Ring with $N = 5$ coupled neurons obeying Dale's Law: each neuron couples in a purely excitatory or purely inhibitory manner to its two neighbors;  solid arrows represent positive,  excitatory connections, and dashed lines negative, inhibitory ones.  The $5 \times 5$ matrix $\boldsymbol G$ corresponding to this particular choices of signs is indicated on the right, where only the sign on the nonzero matrix elements is indicated.    Note that non-zero connections in the same row have the same sign.}
\label{dale}
\end{figure}

Figure~\ref{dale} shows a network with $N=5$ that respects Dale's law, and the signs of the non-zero elements of the corresponding matrix. We now argue that the results presented in this paper are readily extended also to this scenario, which is more realistic for neural networks.
	
To take this situation into account, we consider (taking $g=0$ for now)
\begin{align}\label{eq-h140}
\boldsymbol G=
\begin{pmatrix}
 & \sigma_1 &  &  &  &  &  \\
\sigma_2 &  & \sigma_2 &   &  &  &  \\
 & \sigma_3 &  & \sigma_3 &  &  &  \\
 &  & \sigma_4 &  & \ddots & &  \\
 &  &   &\ddots  &  & \ddots & \\
 &  &  &&\ddots && \sigma_{N-1} \\
 &  &  &  & & \sigma_N & \\
\end{pmatrix}
\end{align}
instead of $\boldsymbol M$ in Eq.~\eqref{eq-h80}, where each of $\sigma_j$ randomly takes $\pm 1$ with probability $1/2$, although similar considerations apply to the more general probability distribution of Eq.~\eqref{GrindEQ__5_}. The value of $\sigma_j$ indicates whether the two connections out of the $j$th neuron are excitatory or inhibitory.

According to the previous argument, the spectrum depends only on the product of opposing off-diagonal elements.
In the case of the matrix~\eqref{eq-h140}, we can regard the $N-1$ quantities $\{\sigma_j\sigma_{j+1}=\pm 1\|j=1,2,\ldots,N-1\}$ as independent random variables, just as for the matrix of Eq.~\eqref{eq-h80} we can regard the $N-1$ quantities $\{s_x^+s_x^-=\pm 1\|x=1,2,\ldots,N-1\}$ as independent.
Therefore, the matrices~\eqref{eq-h80} and~\eqref{eq-h140} are statistically isospectral;
see Fig.~\ref{fig-h3}(b) for the spectrum for one random sample, obeying Dale's law, to be compared with Fig.~\ref{fig-h1}(b).

The statistical isospectrality does not change much when we introduce the boundary terms $\boldsymbol G_{1N}=\sigma_N$ and $\boldsymbol G_{N1}=\sigma_1$, because the perturbation of these terms to the spectrum is of order $1/N$ at most (and exponentially small if the states are localized).
The only difference in the statistics is the fact that the product of all super- and sub-diagonal elements of the matrix~\eqref{eq-h80}, including the boundary terms $\boldsymbol M_{1N}=s_N^+$ and $H_{N1}=s_N^-$, is random and can take $\pm1$, but that of the matrix~\eqref{eq-h140} is always $+1$.

Finally, introduction of the asymmetry parameter $g$ to $\boldsymbol G$ as in
\begin{align}\label{eq-h150}
\boldsymbol G(g)=
\begin{pmatrix}
 & e^{g}\sigma_1 &  &  &  &  &  e^{-g}\sigma_1\\
e^{-g}\sigma_2 &  & e^{g}\sigma_2 &   &  &  &  \\
 & e^{-g}\sigma_3 &  & e^{g}\sigma_3 &  &  &  \\
 &  & e^{-g}\sigma_4 &  & \ddots & &  \\
 &  &   &\ddots  &  & \ddots & \\
 &  &  &&\ddots && e^{g}\sigma_{N-1} \\
e^{g}\sigma_N &  &  &  & & e^{-g}\sigma_N & \\
\end{pmatrix},
\end{align}
has the same effect as we showed in Sec.~\ref{subsec-h3} for the matrix~\eqref{eq-h90};
see Fig.~\ref{fig-h3}(c) for a numerical illustration for one random sample. Note the close similarity with Fig.~\ref{fig-h3}(a), for the matrix $\boldsymbol{M}(g)$, which is unconstrained by Dale's law.

\section {Localization properties}
\label{loc_sect}
We now investigate the localization properties of the model via three different and complimentary routes:
\begin{enumerate}
\renewcommand{\labelenumi}{(\roman{enumi})}
\item By calculating the participation ratio of eigenmodes obtained via exact diagonalization;

\item By using the transfer-matrix approach, and the equivalence between the Lyapunov exponents and the inverse localization length;

\item By numerically calculating the density-of-states (DOS) via exact diagonalization, and inferring the localization length via the Thouless relation~\cite{thouless}, as generalized to localization with complex eigenvalues in Ref.~[\onlinecite{derrida}].
\end{enumerate}

We find analytically that for $g=0$ and for any $u$ that the localization length is infinite at $\lambda=0$ (\textit{i.e}., the inverse localization length $\kappa$ vanishes at the origin), suggesting a diverging localization length as $|\lambda| \rightarrow 0$. Such a divergence is strongly supported by our numerical results. Interestingly, we find that in contrast to the results of Ref.~[\onlinecite{derrida}], the dependence $\kappa$ on $\lambda = x+ i y$  in the vicinity of the origin is not isotropic. Through the Thouless relation, which we elaborate on below, we will show that this property is connected to the vanishing DOS at the origin. In the following, we elaborate on the different methods and compare the results.

\subsection {Localization properties from numerical diagonalization}

\begin{figure}
\includegraphics[width=.43\textwidth]{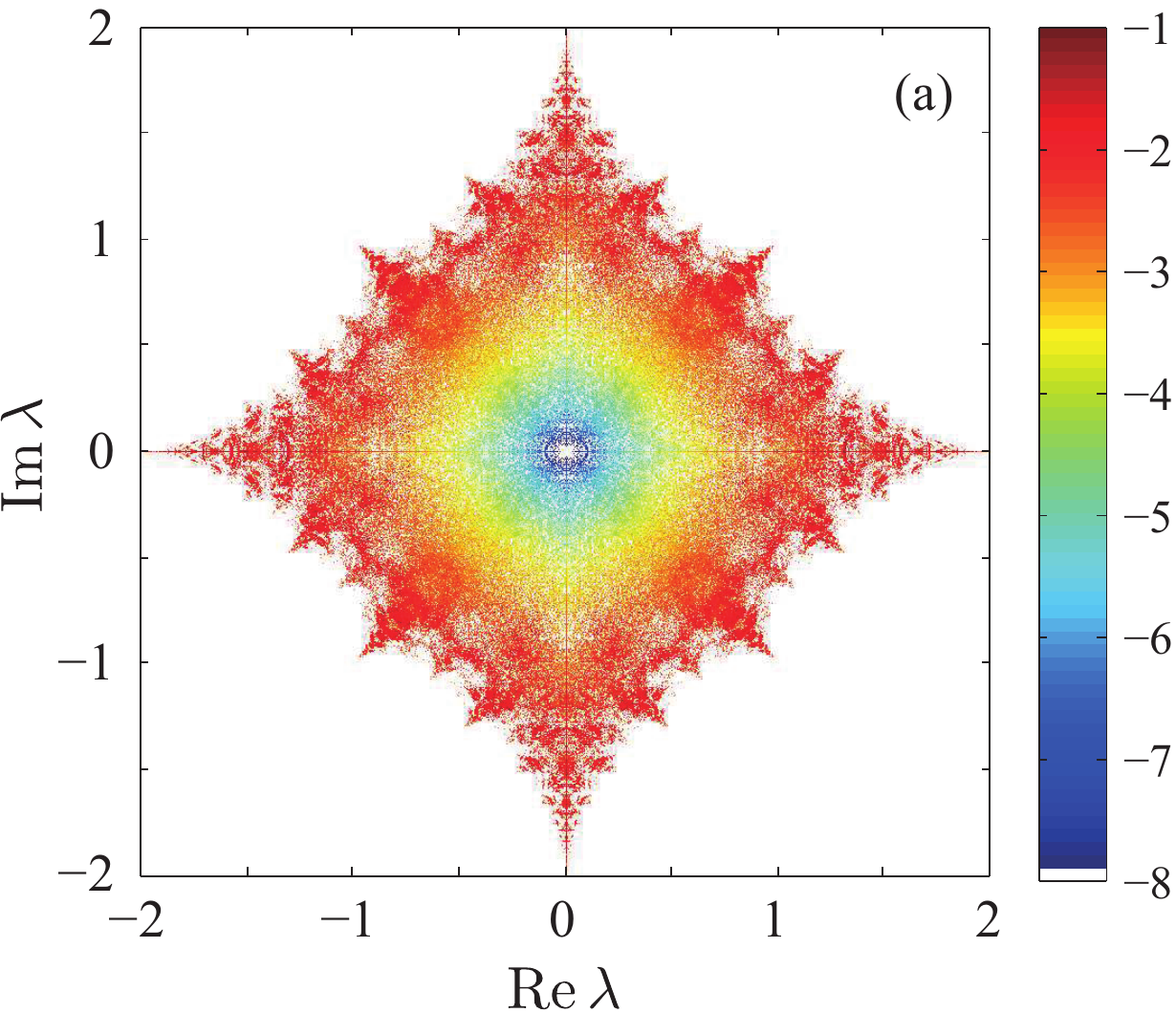}

\vspace{\baselineskip}
\includegraphics[width=.43\textwidth]{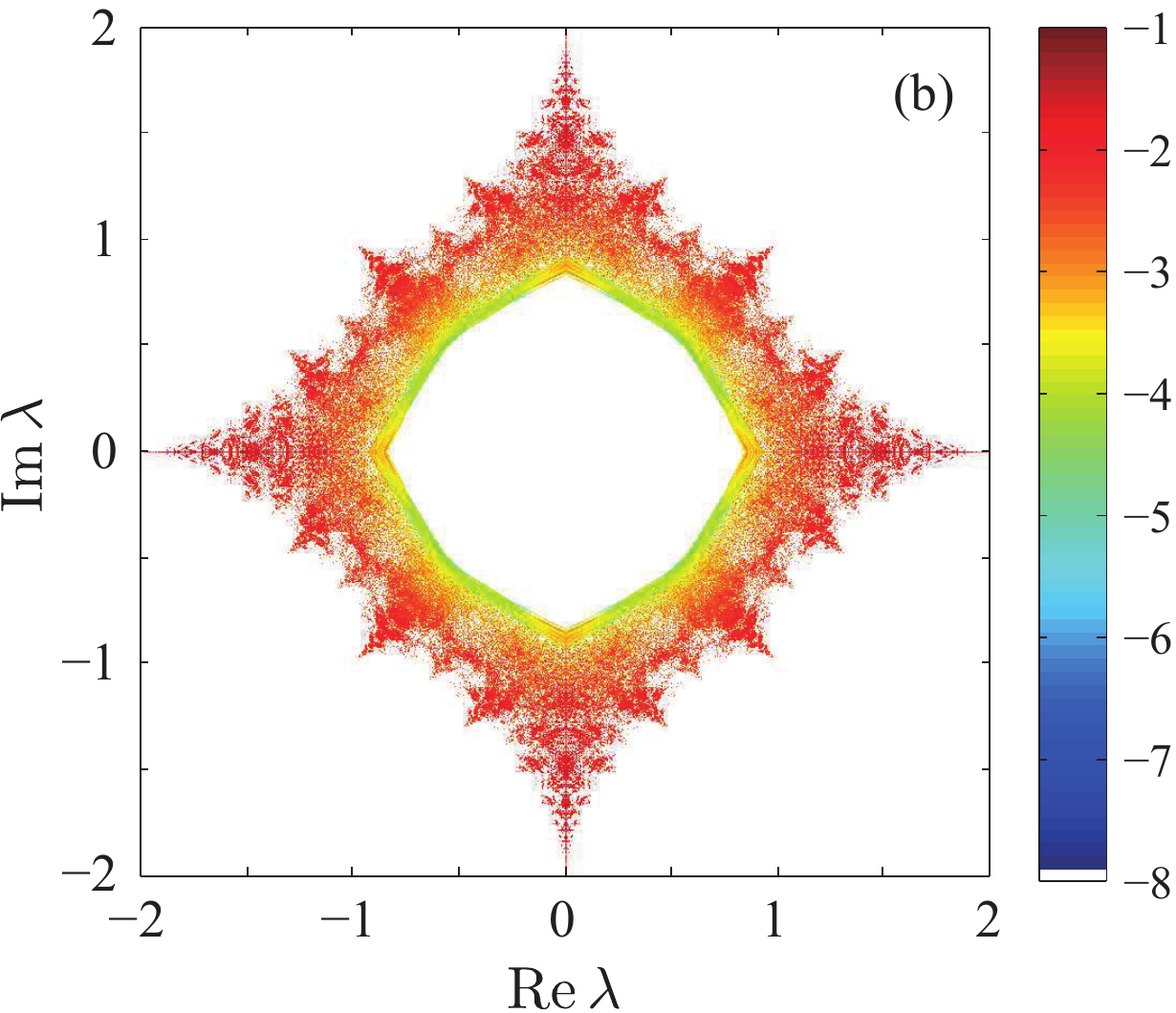}
\caption {(a) Inverse participation ratio (IPR) as a function of eigenvalue, obtained via numerical diagonalization of 10,000 matrices of dimension $5000 \times 5000$ with $g=0$ and periodic boundary conditions. With $g=0$, the results with open boundary conditions would be nearly identical. The color bars indicate the inverse localization length on a logarithmic scale. Note that the background (where there are no states) is white; the fractal nature of the spectrum implies that the IPR is not evaluated everywhere, but only on the support of the DOS. (b) IPR for the same parameters, but with $g=0.1$ and periodic boundary conditions. In this case the states become delocalized on the rim of the hole in the DOS.} \label {ipr_fig}
\label{fig1_}
\end{figure}

A useful measure of the localization of an eigenvector is its participation ratio, defined as
\begin{align}
 P \equiv  \sum_j |\psi_j|^2/ \sum_j |\psi_j|^4.
\end{align}
Indeed, a perfectly localized eigenvector with support only at a single site would have $P=1$, while a perfectly delocalized one (with $\psi_j=1/\sqrt{N}$ for every $j$) has $P=N$.
By averaging the participation ratio, or its inverse, we may gain insights into the localization properties of the system. Figure~\ref{ipr_fig} shows the results of numerical diagonalization of 10,000 matrices of dimension $5000 \times 5000$, performed on Harvard's ``Odyssey'' cluster. These matrices are given by Eq.~\eqref{GrindEQ__4_}, with $f=1/2$ and $u=1$.

Figure~\ref{fig1_}(a) shows the tendency of states to be delocalized near the origin (vanishing inverse localization length (IPR)), becoming more localized away from the origin. However, while this is a direct and straightforward method, in the next subsections~\ref{subsec3-B} and~\ref{subsec3-C}, we will find the localization length more accurately; we will show that while it diverges near the origin, it does not have a radial symmetry, and only achieves radial symmetry away from the origin.

Upon repeating the analysis for $g=0.1$ (Fig.~\ref{fig1_}(b)), we see that the hole in the DOS is accompanied by a diverging localization length on its rim. Later we will show that the model exhibits spectral rigidity: the localized eigenmodes away from the rim of the hole are insensitive to changes in $g$.

We now comment briefly on the effect of periodic boundary \textit{vs}.\ open conditions.  The arguments given in Sec.~\ref{symmetries} suggest (and numerical diagonalizations confirm) that the $g=0$ spectrum is nearly identical when periodic instead of open boundary conditions are employed in Fig.~\ref{fig1_}(a).  In contrast, the hole and extended states in the $g=0.1$ spectrum disappear when periodic boundary conditions are replaced by open boundary conditions in Fig.~\ref{fig1_}(b).   The invariance of spectrum follows from the similarity transformation leading to Eq.~\eqref{eqA} of Sec.~\ref{symmetries}, after taking the limits $s_{N}^{+} \to 0,{\rm \; }s_{N}^{-} \to 0$, which breaks the chain.    Nevertheless the hole has a physical interpretation even for open chains:  Although all eigenvalues retain their $g=0$ values, eigenfunctions inside the hole become edge states, piled up on one side of the broken chain.

\subsection {Transfer matrix approach}
\label{subsec3-B}

A well-established method for finding the localization length of a one-dimensional system calculates the Lyapunov exponent via the transfer matrix technique~\cite{derrida}. If $\psi_n$ is the eigenfunction amplitude on the $n$th site, the $2 \times 2$ transfer matrix connecting the vector
$
\begin{pmatrix}                                                                                                                                                                                                                                             \psi_n \\                                                                                                                                                                                                                                             \psi_{n+1} \\                                                                                                                                                                                                                                           \end{pmatrix}
$
to the vector
$
\begin{pmatrix}                                                                                                                                                                                                                                             \psi_{n-1} \\                                                                                                                                                                                                                                             \psi_{n} \\                                                                                                                                                                                                                                           \end{pmatrix}
$
with eigenvalue $\lambda$ is given by
\begin{align}
 \boldsymbol T_n =
          \begin{pmatrix}
            0 & 1 \\
            -e^{-2g} s^{-}_{n-1}/s^+_n & \lambda e^{-g} /s^+_n \\
          \end{pmatrix}, \label{transfer}
\end{align}
where we do not include diagonal disorder and $s^{-}_{n-1}$ and $s^+_n$ are independent random variables representing the off-diagonal randomness.

The Lyapunov exponent can be extracted by taking the limit
\begin{align}
 \kappa \equiv \lim_{N \rightarrow \infty} \langle \log (|| \boldsymbol T_N \cdot \boldsymbol T_{N-1} ... \boldsymbol T_1 ||)\rangle /N ,
\end{align}
where $||..||$ denotes the norm of the matrix, and $\langle..\rangle$ ensemble averaging over the quenched disorder. It can be proven that under quite general conditions the limit exists, and $\kappa$ equals the inverse of the localization length~\cite{ishii}, which we identify (up to constants of order unity) with the inverse participation ratio of Sec.~\ref{loc_sect}A.

This procedure provides a numerically attractive route to finding the localization length, without having to diagonalize large matrices. However, in practice $N$ has to be large in order for the method to be accurate, which implies that the product will result in a matrix with a large norm, imposing computational difficulties. We resolved this problem by working with the recursive relation for the quantity $r_n \equiv \psi_{n+1}/\psi_n$ (note that unlike Ref.~[\onlinecite{derrida}], in our definition $r$ is a complex number). From Eq.~\eqref{transfer} we immediately find that
\begin{align}
 r_{n+1} = -(s^{-}_{n-1}/s^+_n) e^{-2g}/r_n + \lambda e^{-g}/s^+_n .\label{riccati1}
\end{align}

In this case, the values of $r_n$ are well-behaved also for large $n$, leading to robust numerics. Upon evaluation of $r_1...r_N= \psi_{n+1}/\psi_1$, the Lyapunov exponent can be found in a similar fashion as
\begin{align}
 \kappa(\lambda) = \frac{1}{N} \sum_{j=1}^N {\log{|r_j|}} . \label{riccati2}
\end{align}
It is beneficial to omit the values of $r_j$ at the beginning of the sequence, to reduce the effects of the initial conditions, though in the limit of large $N$ this is not strictly necessary.

\begin{figure}
\includegraphics[width=.43\textwidth]{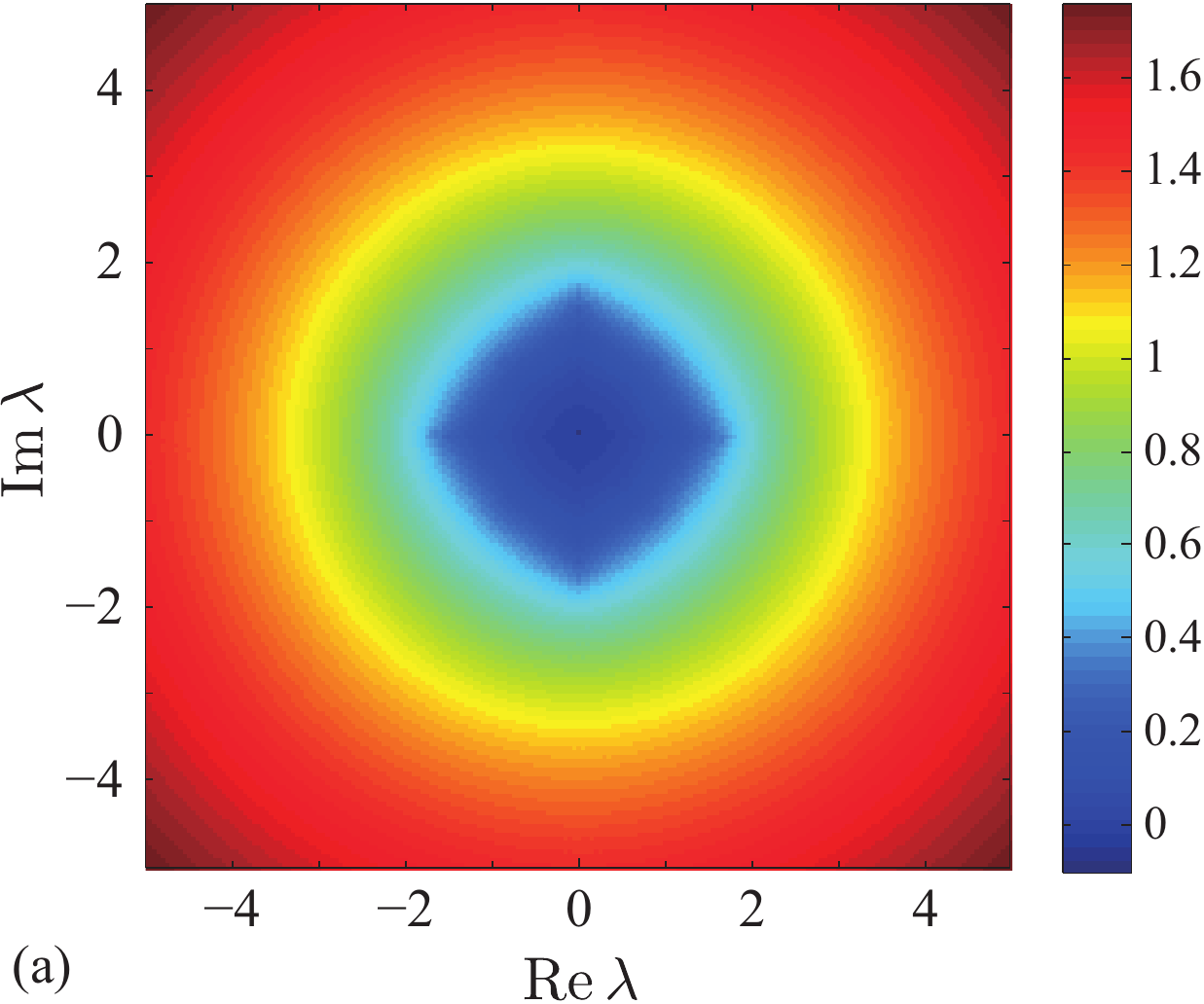}\\

\vspace{\baselineskip}
\includegraphics[width=.43\textwidth]{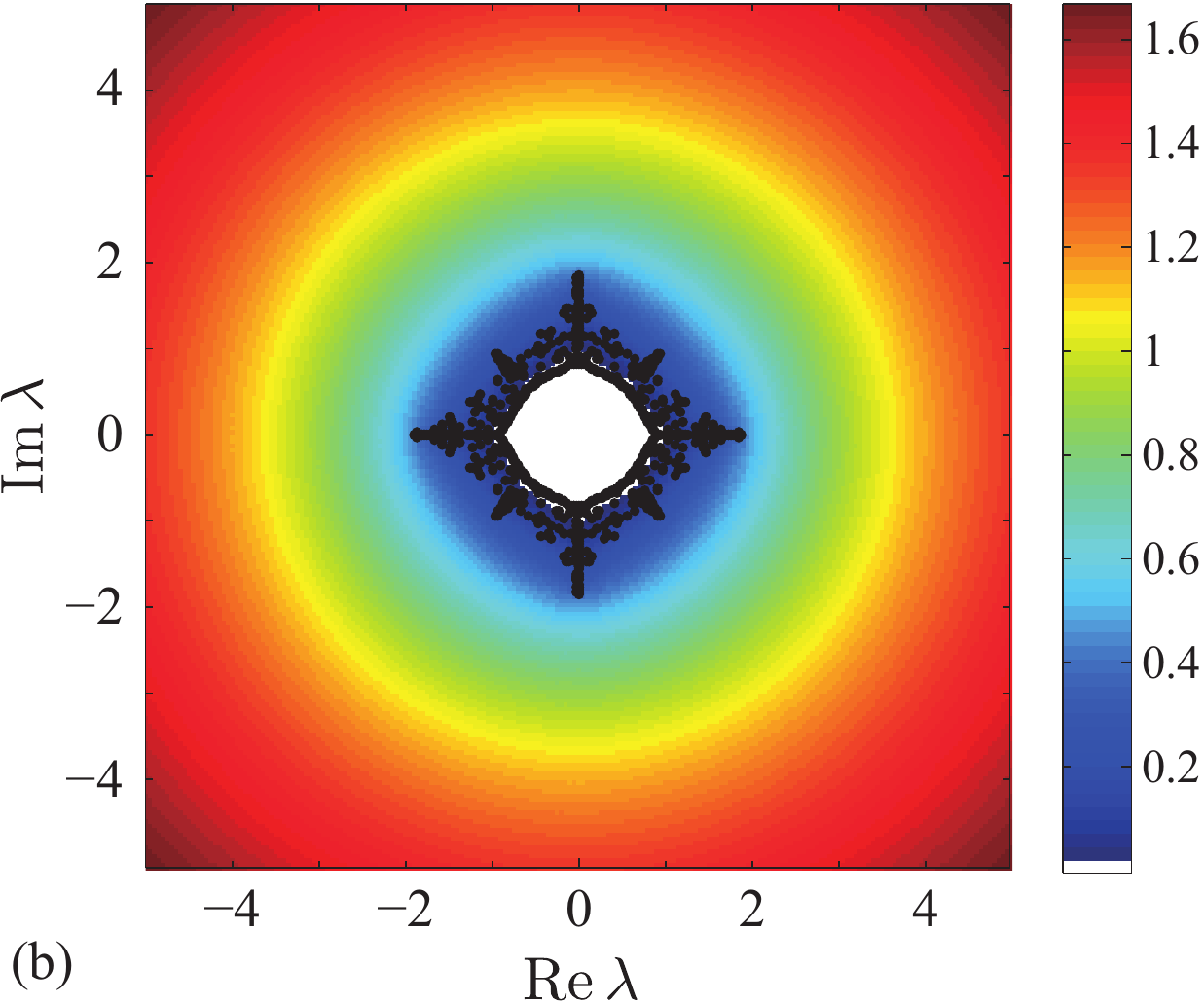}\\
\caption {(a) Inverse localization length $\kappa(\lambda)$ as a function of eigenvalue, obtained via Eq.~\eqref{riccati2} and the recursion equation~\eqref{riccati1}. Each point is obtained via 10,000 iterations. In this case $\kappa$ does not take negative values anywhere. (b) The colored  map shows $\kappa(\lambda)$ for the same parameters, but with $g=0.1$. There is a finite region in the vicinity of the origin with \emph{negative} values of $\kappa$, which was given a white color (not included in the color map). This region corresponds to the gap of Fig.~\ref{ipr_fig}(b); the black dots superimposed on the plot are the result of the diagonalization of a single $1000 \times 1000$ matrix with $g=0.1$. Note that the boundary of the white hole corresponds almost exactly to the rim of the extended states found via the exact numerical diagonalization. } \label {riccati_fig}
\end{figure}

Using this method, we obtained Fig.~\ref{riccati_fig}, which corroborates and complements the results of the exact numerical diagonalization. While Fig.~\ref{fig1_} is computationally expensive, generating Fig.~\ref{riccati_fig} takes several minutes on a PC, a testimony to the power of this technique. Note, however, that Eqs.~\eqref{riccati1} and~\eqref{riccati2} \emph{always} deliver a value for $\kappa(\lambda)$, regardless of whether there is actually a normalizable eigenfunction at that particular value of $\lambda$.

\subsection {Connection to the density-of-states via the Thouless relation}
\label{subsec3-C}

A classic result in the theory of Anderson localization in one dimension is an elegant relation connecting the density-of-states to the localization length, due to Thouless~\cite{thouless}. This relation can readily be generalized to the non-Hermitian case~\cite{derrida}, where it states that
\begin{align}
 \nabla^2 \kappa(x,y) = \rho(x,y) . \label{thouless_direct}
\end{align}
Here, the complex eigenvalue is $\lambda = x+iy$, and $\rho(x,y)$ is the density-of-states. This equation can be inverted, using the well-known analogy with 2d electrostatics, whereby $\rho(x,y)$ represents a collection of infinite, charged wires, perpendicular to the complex plane, each associated with a logarithmic potential. Therefore we have
\begin{align}
 \kappa(x,y) = \int \rho(x',y') \log(|r-r'|) dx dy + C . \label{thouless_inv}
\end{align}
In the case $g=0$, the constant is given by~\cite{molinari}
\begin{align}
 C = \langle \log(|s^+_j|) \rangle =  \langle \log(|s^-_j|) \rangle ,
\end{align}
\textit{i.e}., the average of the logarithm of the random matrix elements. Hence, in the case we are focusing on where $|s^+_j|=1$, we find that $C=0$. In the next section we shall show how the results for $\kappa(x,y)$ for finite $g$ can be mapped to the $g=0$ behavior, which will show that in the more general case we have
\begin{align}
 C = \frac{1}{2}\langle \log(|s^-_j/s^+_j|)\rangle = -g,
\end{align}
which follows from Eq.~\eqref{GrindEQ__8_}.

\begin{figure}
\includegraphics[width=.43\textwidth]{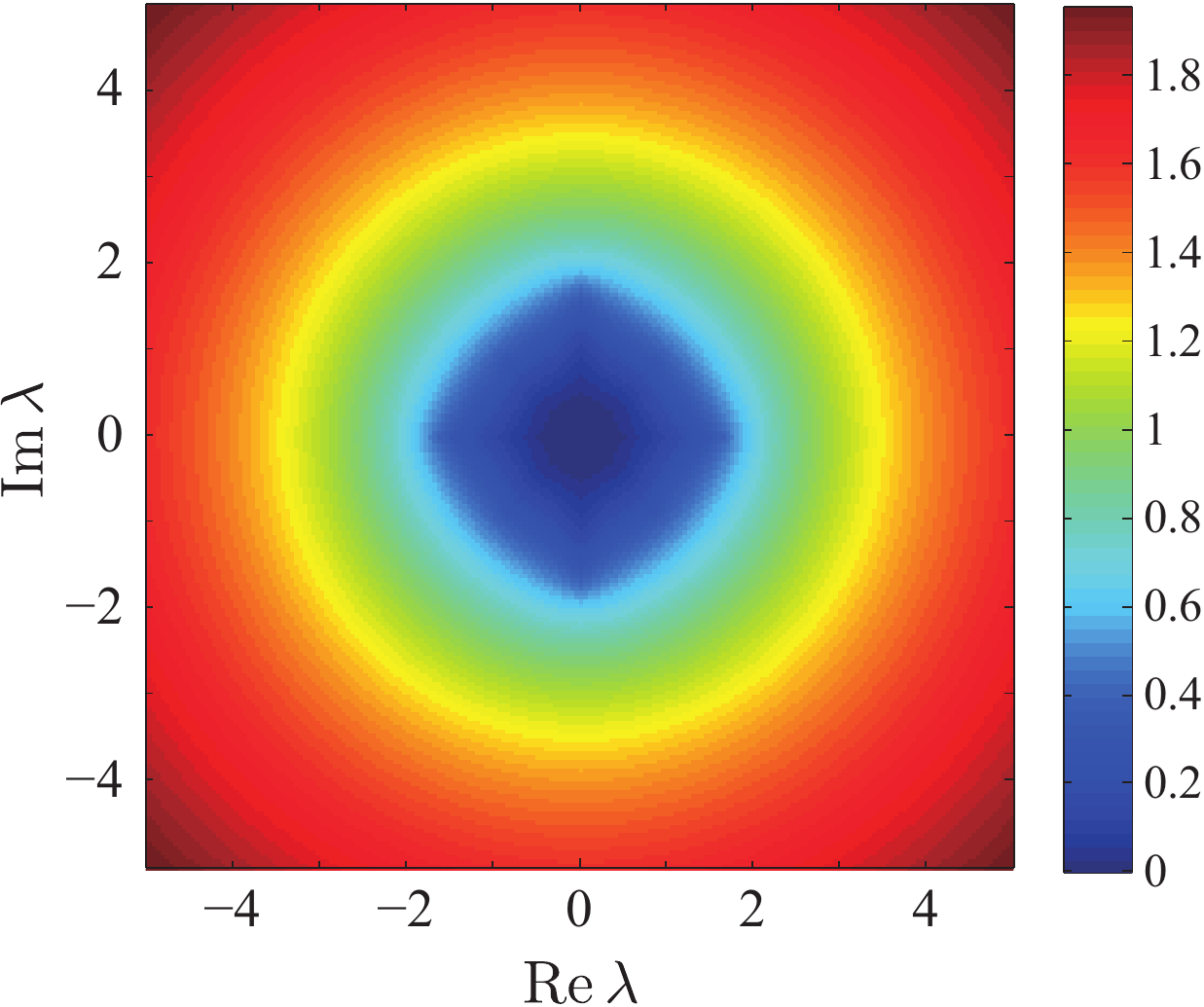}\\
\caption {The Lyapunov exponent $\kappa(\lambda)$ was extracted from the diagonalization of a single $1000 \times 1000$ matrix with $g=0$, using the electrostatic relation~\eqref{thouless_inv}. The result is very similar to Fig.~\ref{riccati_fig}(a), which was obtained via the Ricatti recursion relations.} \label {thouless1}
\end{figure}

This remarkable relation allows us to go back-and-forth between the two very different numerical procedures: obtaining $\kappa$ via the recursion relation and obtaining $\rho$ via exact numerical diagonalization. Indeed, using a \emph{single} realization of $1000 \times 1000$ matrix and applying this formula allows us to recover the Lyapunov exponent dependence on energy, shown in Fig.~\ref{thouless1} for the case $g=0$.

\subsection {Hole in spectrum corresponds to contours of Lyapunov exponent}
\label{hole}

Consider the recursion relation of Eq.~\eqref{riccati1}. It is easy to ``gauge away'' the effect of $g$, by making the transformation
\begin{align}
 y_n \equiv r_n e^{g} ,
\end{align}
upon which the equation takes the form
\begin{align}
 y_{n+1} = -\left(\frac{s^-_{n-1}}{s^+_n}\right)/y_n +\lambda/s^+_n .\label{riccati_v2}
\end{align}
This representation implies that for any complex eigenvalue $\lambda = x+ i y$, the effect of $g$ is to decrease the Lyapunov exponent by an amount $g$:
\begin{align} \label{GrindEQ__8_}
\kappa (x ,y ;g)=\kappa (x ,y ;0)-g
\end{align}
Hence, consistent with the gauge transformation result of Eq.~\eqref{eq-h130}, for any $g>0$ all states which previously had $\kappa<g$ will acquire a \emph{negative} Lyapunov exponent. Since all states must be normalizable, the region with negative $\kappa$ will not support any eigenfunctions, and corresponds to the ``hole'' or gap seen in Figs.~\ref{dos} and~\ref{fig-h3}. This argument implies that the hole boundary corresponds to contour where $\kappa=g$, consistent with Fig.~\ref{riccati_fig}(b), where the results of exact diagonalization are superimposed on top of the Lyapunov exponent heatmap.

\begin{figure}
\includegraphics[width=.37\textwidth]{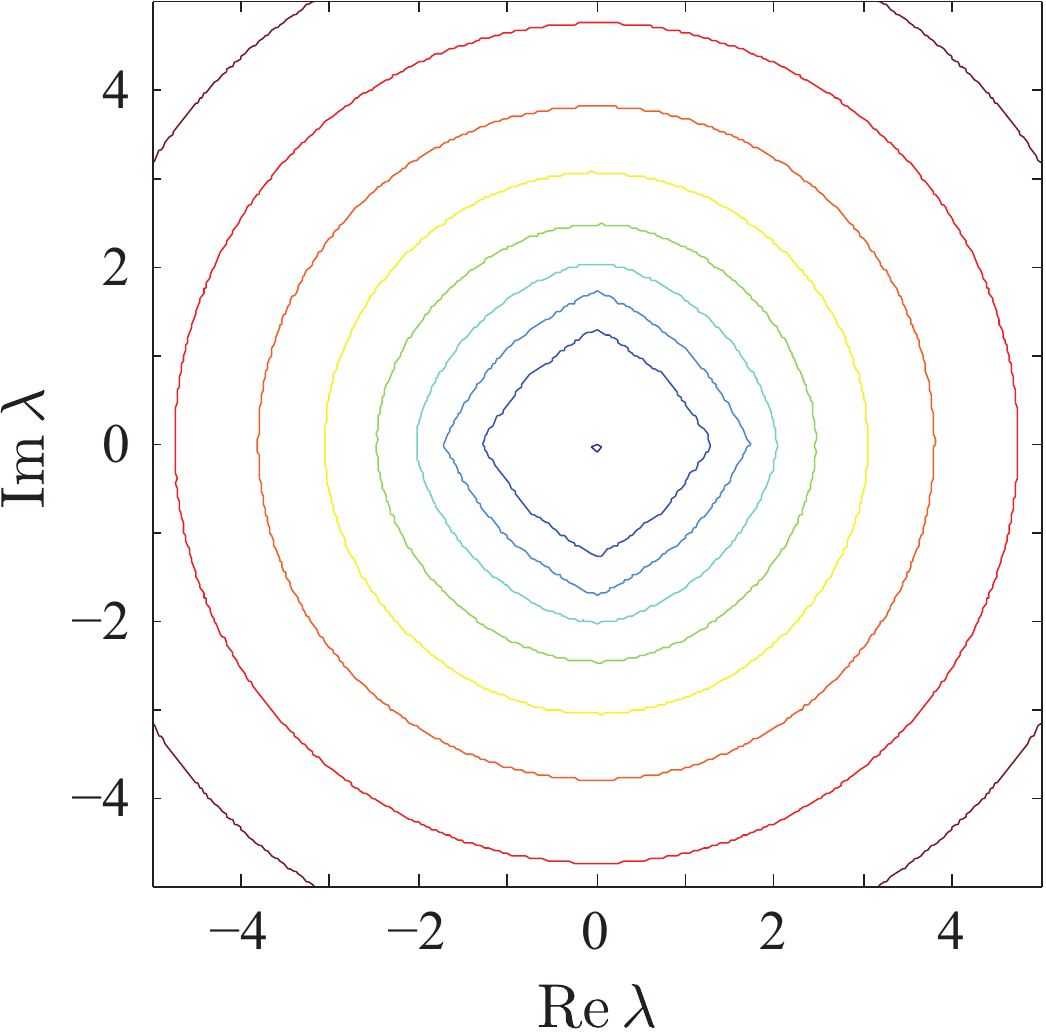}\\

\caption {Constant $\kappa$ contours of the heatmap of Fig.~\ref{riccati_fig}(a). } \label {contour}
\end{figure}


In Fig.~\ref{contour}, the contours of constant $\kappa$ are shown. As expected from the electrostatic Thouless relation, away from the effective support of the DOS the contours become circular, $\kappa(x,y) \propto \log(x^2+y^2)$, since all ``charges'' associated with the complex eigenvalues act as if they were concentrated at the origin. Close to the origin the contours obtain a roughly diamond or octagonal shape. This behavior is consistent with the vanishing DOS suggested by Fig.~\ref{dos}(a); via the Thouless relation, Eq.~\eqref{thouless_direct}, if $\kappa$ had a perturbative expansion such as $\kappa \sim x^2+y^2$ (see Ref.~[\onlinecite{derrida}] for such result in a related model), then the DOS at the origin would have a finite, non-vanishing value.

\begin{figure}
\includegraphics[width=.48\textwidth]{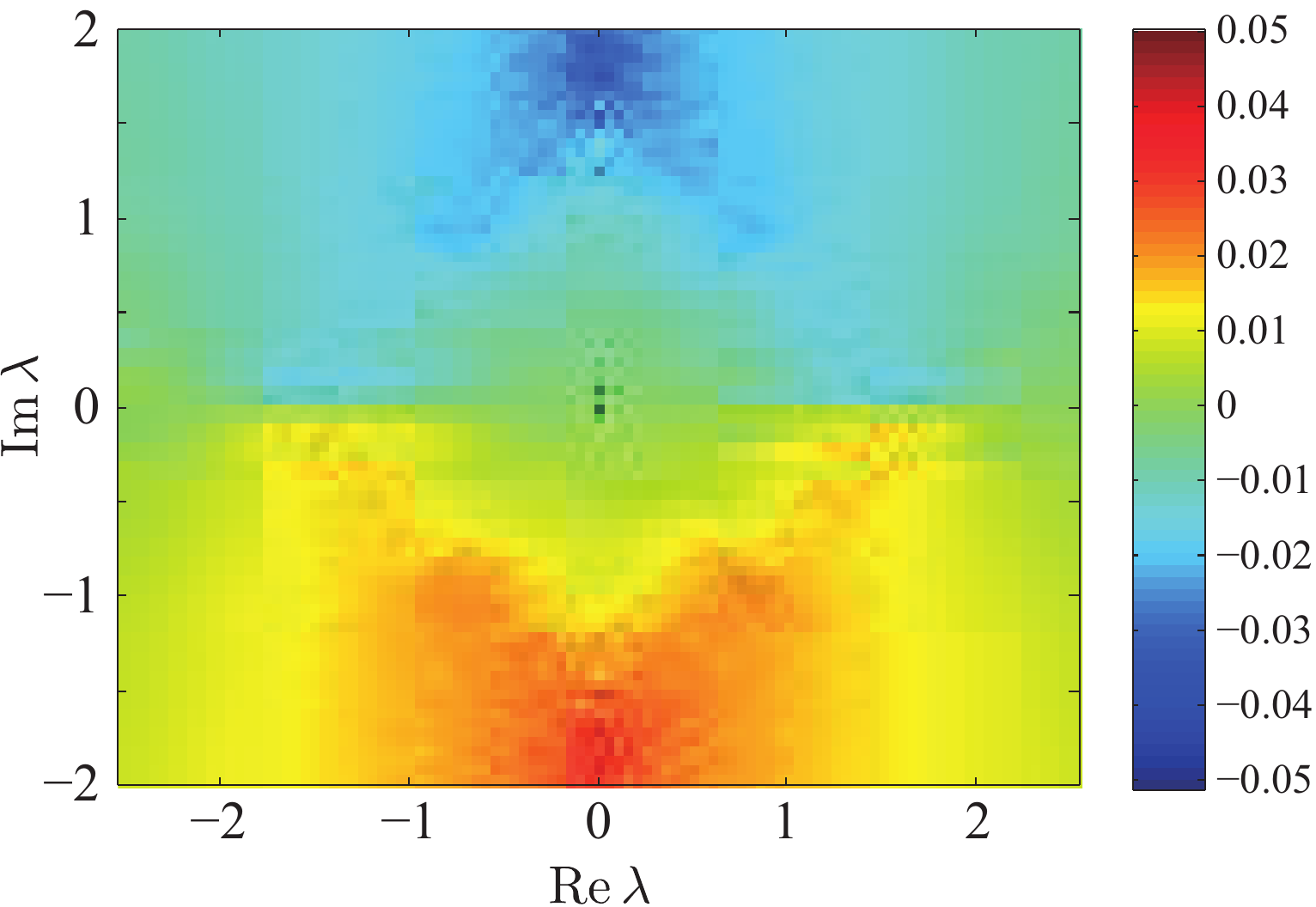}\\
\caption {The $y$ component of the gradient of the Lyapunov exponent. The abrupt change along the $x$ is consistent with a condensation of eigenvalues along the real axis. } \label {gradient_y_riccati}
\end{figure}

\begin{figure}
\includegraphics[width=.37\textwidth]{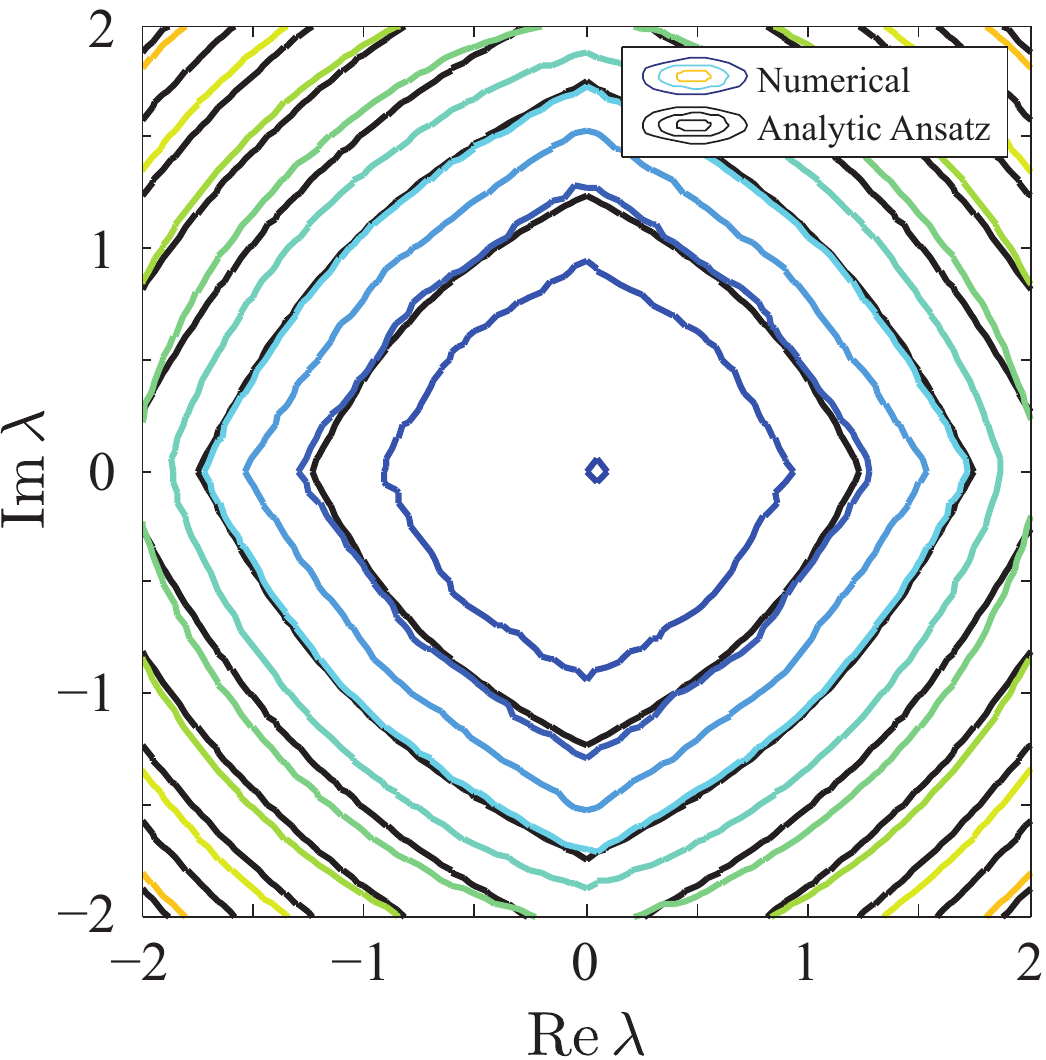}\\
\caption {Numerically extracted contours of constant Lyapunov exponent near the origin, compared with those of Eq.~\eqref{ansatz_eq}.} \label {ansatz_fig}
\end{figure}

Furthermore,  Fig.~\ref{gradient_y_riccati} shows the results for the $y$ component of the \emph{gradient} of the Lyapunov exponent, suggesting a $\delta$-function contribution to the DOS along the $x$ axis. Similar results can be obtained for the $y$ axis. For $u=1$, the strength of the singular DOS along the $x$ and $y$ axis decays \emph{linearly} close to the origin, as shown in Fig.~\ref{dos2}.
These results lead us to the following ansatz for the behavior of $\kappa$ near the origin for $u=1$:
\begin{align}
 \kappa(x,y) \sim (|x|+|y|)\sqrt{x^2+y^2} , \label{ansatz_eq}
\end{align}
\textit{i.e}., it is a product of $L_1$ and $L_2$ norms. This ansatz is consistent with the eigenvalue condensations onto the $x$ and $y$ axis, and their linear density shown in Fig.~\ref{dos2}.
When appropriate higher-order cubic terms are added to the ansatz of Eq.~\eqref{ansatz_eq}, the function becomes harmonic away from the $x$ and $y$ axes (\textit{e.g}.: by replacing $d=\sqrt{x^2+y^2}$ with $[1-e^{-2d}]$), consistent with the vanishing DOS at the origin. The good agreement of the equipotential contours of the ansatz of Eq.~\eqref{ansatz_eq} and of the numerically evaluated Lyapunov exponent is shown in Fig.~\ref{ansatz_fig}.

\subsection {Vanishing of the Lyapunov exponent at the origin}
\label{vanish}
It is easy to see that for \emph{any} distribution of the hopping matrix elements, the Lyapunov exponent must vanish at the origin (in contrast to the behavior of models with additional diagonal disorder~\cite{molinari}). To see this, consider the transfer matrix Eq.~\eqref{transfer} for $\lambda =0$. The product of two adjacent transfer matrices is in this case diagonal:
\begin{align}
 \boldsymbol S_n = \boldsymbol T_n \boldsymbol T_{n-1} = e^{-2g}
          \begin{pmatrix}
            -\frac{s^-_{n-1}}{s^+_n} & 0 \\
            0 & -\frac{s^-_{n-2}}{s^+_{n-1}}\\
          \end{pmatrix}
\end{align}
with the elements on the diagonal being the \emph{ratio} of two random variables. Therefore the Lyapunov exponent is given by
\begin{align}
 \kappa =\frac{1}{2}\langle \log(|s^-_j/s^+_j|)\rangle =0 ,
\end{align}
a result that holds provided that $s^-_j$ and $s^+_{j+1}$ are chosen from identical, independent probability distribution functions, as in Eq.~\eqref{GrindEQ__5_}.
The vanishing value of $\kappa$ at the origin is numerically corroborated in Fig.~\ref{riccati_fig}. In fact, for $\lambda =0$ (and for any probability distribution function $P(s)$), there is an extended eigenfunction of Eq.~\eqref{GrindEQ__4_} that reads
\begin{align} \label{GrindEQ__7_}
&{\left| \lambda =0,g=0 \right\rangle} \nonumber\\
&={\left| 1 \right\rangle} +\sum _{m=1}^{(N-1)/2}(-1)^{m}
\frac{s_{1}^{-} s_{3}^{-} ...s_{2m-1}^{-} }{s_{2}^{+} s_{4}^{+} ...s_{2m}^{+} }  {\left| 2m+1 \right\rangle}  \nonumber\\
&\equiv {\left| 1 \right\rangle} +\sum _{m=1}^{(N-1)/2}\psi _{m} {\left| 2m+1 \right\rangle}, \nonumber\\
&\psi_{m} =(-1)^{m} \frac{s_{1}^{-} s_{3}^{-} ...s_{2m-1}^{-} }{s_{2}^{+} s_{4}^{+} ...s_{2m}^{+} }
\end{align}
where we have assumed $N$ is odd and the amplitudes on all even sites vanish.   A similar state can be constructed for an even number of sites, with a mild restriction on $s_{N}^{+} $ and $s_{N}^{-} $ in both cases when periodic boundary conditions are imposed.   That this zero energy state is indeed extended follows from the definition of the inverse localization length within the transfer matrix method,[26-29] $\kappa =\mathop{\lim }\limits_{N\to \infty } \frac{1}{N} \left\langle \log \left|\psi _{N} \right|\right\rangle $, where the average is over the probability distributions of the matrix elements in Eq.~\eqref{GrindEQ__7_}.

\subsection {Spectral rigidity outside the gap}

Consider the model for $g=0$, and ``ramp up'' $g$. As we argued in Sec.~\ref{hole} and as was discussed in Ref.~[\onlinecite{molinari}] in the context of a related model, this results in a hole that tracks the contours of constant Lyapunov exponent. Thus, as $g$ increases, the hole widens and ``sweeps away'' the eigenvalues in its vicinity. The hole hence acquires a finite fraction of the spectrum, concentrated on its one-dimensional rim. Since the rim of the hole corresponds to diverging Lyapunov exponent, these states have all become \emph{delocalized} by the finite value of $g$, while the states outside the hole are still localized, as explained in Sec.~\ref{subsec-h3}. These states are insensitive to the boundary conditions, and their eigenvalues will not be modified by  $g$. 

\begin{figure}
\includegraphics[width=.37\textwidth]{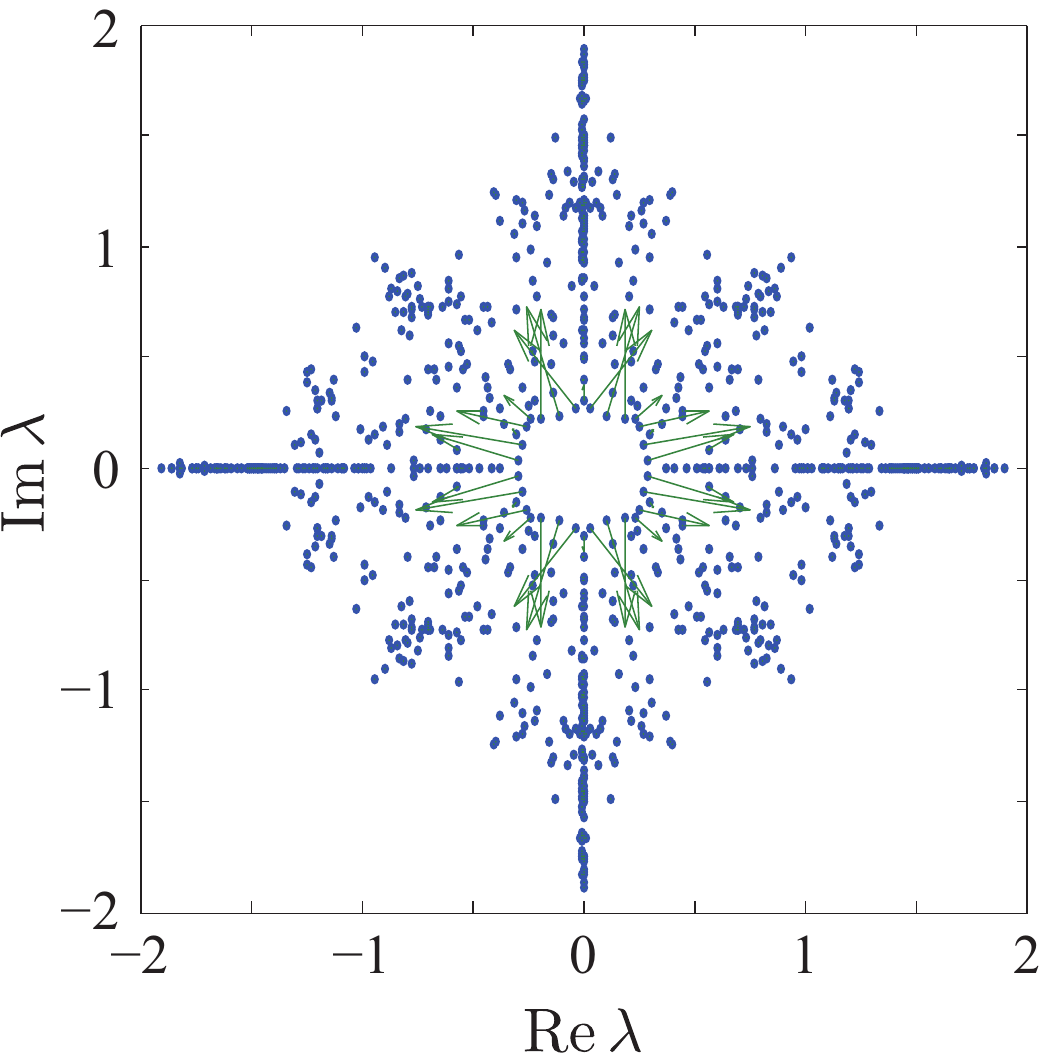}

\caption {The eigenvalues of an exact numerical diagonalization of a matrix with $N=1000$, $g=0.01$ (blue dots). From each point a line emanates, in the direction of the eigenvalue velocity in the complex plane (\textit{i.e}.\ $\frac{d\lambda}{dg}$), and with length proportional to the velocity. For eigenvalues away from the rim of the hole, no line is visible; spectral rigidity implies vanishing velocities in this regime. } \label {spectral}
\end{figure}

This spectral rigidity is illustrated in Fig.~\ref{spectral}. To calculate the eigenvalue velocity $d\lambda/dg$, we used first-order perturbation theory, which states that this derivative is given by
\begin{align}
 \frac{d\lambda}{dg} = \frac{dx}{dg}+i \frac{dy}{dg}=(\vec{v}^L_\lambda , \boldsymbol B  \vec{v}^R_\lambda) ,
\end{align}
where $\vec{v}^R_\lambda>$ and $\vec{v}^L_\lambda$ are the right and left eigenvectors respectively of the non-Hermitian matrix $\boldsymbol M(g)$ , $(..,...)$ is the scalar product, and the matrix $\boldsymbol B$ is the matrix derivative of the matrix $\boldsymbol M(g)$ with respect to $g$.

\section {Perturbation theory for large $g$}
\label{perturb_sect}
Our problem simplifies for large $g$. In this limit we first neglect all terms of order $e^{-g}$, and the remaining matrix, with periodic boundary conditions, is of the form (illustrated for $N=4$)
\begin{align}
\boldsymbol M=
  \begin{pmatrix}
    0 & s^+_1 e^{g} & 0 & 0 \\
     0 & 0 & s^+_2 e^{g} & 0 \\
    0 &  0 & 0 & s^+_3 e^{g} \\
    s^+_4 e^{g} & 0 & 0 & 0 \\
  \end{pmatrix}
\end{align}
with $s^+_j=\pm 1$ for $f=1/2$ and $u=1$.
We can attempt to ``gauge out'' the signs of $\{s^+_j\}$ by applying a similarity transformation $H=Q^{-1}MQ$ with
\begin{align}
\boldsymbol Q=
  \begin{pmatrix}
    c_1 & 0 & 0 & 0 \\
    0 & c_2 & 0 & 0 \\
    0 & 0 & c_3 & 0 \\
    0 & 0 & 0 & c_4 \\
  \end{pmatrix}.
\end{align}
Choosing $c_1 c_2 = s^+_1$, $c_2 c_3 = s^+_3$ ... $c_n c_1 = s^+_n$ (with each $c_i = \pm 1$) results in a matrix of the form
\begin{align}
\boldsymbol H=e^{g}
  \begin{pmatrix}
    0 & 1 & 0 & 0 \\
     0 & 0 & 1 & 0 \\
    0 &  0 & 0 & 1 \\
    1 & 0 & 0 & 0 \\
  \end{pmatrix}.
\end{align}
Note that $\boldsymbol H$ is proportional to the translational operator for a clockwise rotation of one lattice constant around the ring. This procedure can only be applied when the product of the odd elements $s^+_1 s^+_3...$ equals that of the even elements $s^+_2 s^+_4...$ (which occurs with probability $1/2$). If this is not the case, however, a similar approach can still be pursued (with purely imaginary value of $\{c_i\}$ in this case) leading to similar results.

This matrix is readily diagonalized by plane waves, \textit{i.e}., right eigenvectors  $v^R_j = e^{i k j}$, where the periodic boundary conditions imply that the allowed values of $k$ must be $k= 2 \pi n/ N$, $n=0,1...(N-1)$; note that the left eigenvectors are given by $v^L_j = e^{-i k j}$. The resulting eigenvalues are then
\begin{align}
 \lambda_k = e^{g+i k} ,
\end{align}
\textit{i.e}., except for their magnitude $e^g$ they are the $N$ roots of unity. The eigenvectors of the original matrix $\boldsymbol M$ are plane waves modulated by random sign changes determined by the elements ${s^+_j}$.

So far we concluded that to zeroth order the eigenvalues will sit at \emph{regular intervals on a circle}. We may now introduce the terms with the factor $e^{-g}$ as a perturbation, and calculate the shift of the eigenvalues to the first order in perturbation theory.
The perturbation matrix is of the form (both before and after the similarity transformation)
\begin{align}
\boldsymbol B=e^{-g}
  \begin{pmatrix}
    0 & 0 & 0 & s^-_4 \\
    s^-_1 & 0 & 0 & 0 \\
    0 & s^-_2 & 0 & 0 \\
    0 & 0 & s^-_3 & 0 \\
  \end{pmatrix}.
\end{align}
Within first-order perturbation theory the shift in the $k$th eigenvalue is
\begin{align}
 \delta \lambda_k = (\vec{v}^L_k, \boldsymbol B \vec{v}^R_k) ,
\end{align}
and upon inserting the plane-wave eigenfunctions we have
\begin{align}
 \delta \lambda_k = e^{-g}[ e^{-i k} (s^-_1+s^-_2...+s^-_{n-1}+s^-_N)/N] ,
\end{align}

\begin{figure}
\includegraphics[width=.45\textwidth]{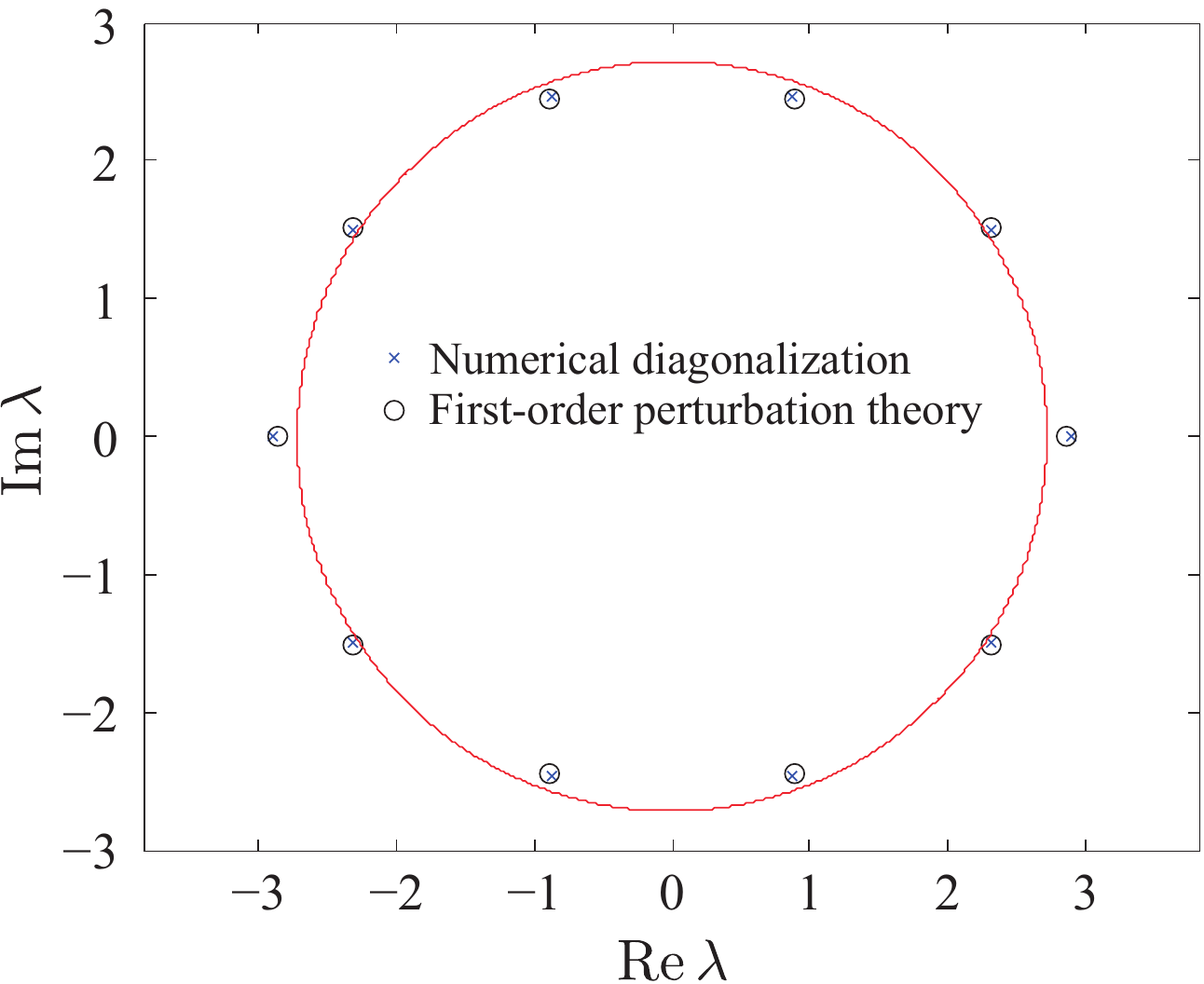}\\

\caption {Comparison of first-order perturbation theory and exact numerical diagonalization, for $g=1$, $N=10$. The red circle has radius $e^{g}$. } \label {perturb}
\end{figure}

Upon invoking the central-limit theorem, for large $N$ we can replace the sum by a Gaussian variable with variance $N$. Hence the eigenvalue will be shifted in the direction $\pm 1/\lambda$, with a magnitude of order $e^{-g}/\sqrt{N}$.
Note that the magnitude of the shift is identical for all eigenvalues. These results are illustrated in Fig.~\ref{perturb}.

It is straightforward to repeat these calculations for hopping elements $s_{j}^{+} $ and $s_{j}^{-} $governed for the more general probability distribution of Eq.~\eqref{GrindEQ__5_}.   After a similarity transformation, $\boldsymbol M\to \boldsymbol M'=\boldsymbol P^{-1} \boldsymbol M\boldsymbol P$, with $\boldsymbol P=\mathop{\mathrm{diag}}\{ 1,{\rm \; }1/s_{1}^{+} ,{\rm \; }1/(s_{1}^{+} s_{2}^{+} ),...\} $ and up to corner matrix elements that do not affect our results as $N\to \infty $, we have
\begin{align}
{\boldsymbol  M'} & {=} {-\sum _{j=1}^{N}\left[e^{g} {\left| j+1 \right\rangle} {\left\langle j \right|} {\rm \; +\; s}_{j}^{+} {\rm s}_{j}^{-} e^{-g} {\left| j \right\rangle} {\left\langle j+1 \right|} \right] {\rm \; \; }} \nonumber\\
&  {\equiv \boldsymbol H+\boldsymbol B},
\end{align}
 where the periodic boundary conditions imply ${\left| j+N \right\rangle} ={\left| j \right\rangle}$.
We recover the plane-wave eigenvectors discussed above for $H$, and find from first-order perturbation theory
\begin{align}
 \left\langle \lambda _{k} \right\rangle =e^{g+ik} +e^{-g-ik} \frac{1}{N} \sum _{j=1}^{N}s_{j}^{+}  s_{j}^{-}.
\end{align}
With the help of the probability distribution of Eq.~\eqref{GrindEQ__5_}, we can now carry out a disorder average, with the result
\begin{align}
 {\lambda _{k} }  {=}  {e^{g+ik} +e^{-g-ik} (1+u)^{2} (f-1/2)^{2} } \nonumber\\
 \equiv e^{g+ik}+\alpha e^{-g-ik}, \alpha = (1+u)^2(f-1/2)^2,
   \label{pert2}
\end{align}
so that the eigenvalues will lie on an ellipse with major axis $e^{g} +\alpha e^{-g}$ and minor axis $e^{g} -\alpha e^{-g} $.   It is straightforward to show for this generalized problem that the fluctuation of the $k$th eigenvalue about its mean values is $O(e^{-g} /\sqrt{N} )h(u,f)$, where $h(u,f)$ is a dimensionless function of order unity.

In Appendix~\ref{second_order}, we go to \emph{second}-order in perturbation theory, and show that it leads to a similar picture qualitatively.

\section {Summary and outlook}
\label{sum_sect}

In this paper we studied a coarse-grained, simplified model for the dynamics of neural networks, which upon linearization close to a steady state leads to the study of the eigenvector spectrum of an ensemble of sparse, non-Hermitian matrices. In contrast to most previous studies in this context, here the connections were only between neighboring neurons, \textit{i.e}., the model included a spatial structure. For concreteness and simplicity, we focused on a ring topology, which is realized in several instances in neuroscience~\cite{ring1, ring2}. An additional parameter in our model, $g$, controlled the directional bias in the neural network, \textit{i.e}., favoring clockwise over counterclockwise connections.

\begin{figure}
\includegraphics[width=.43\textwidth]{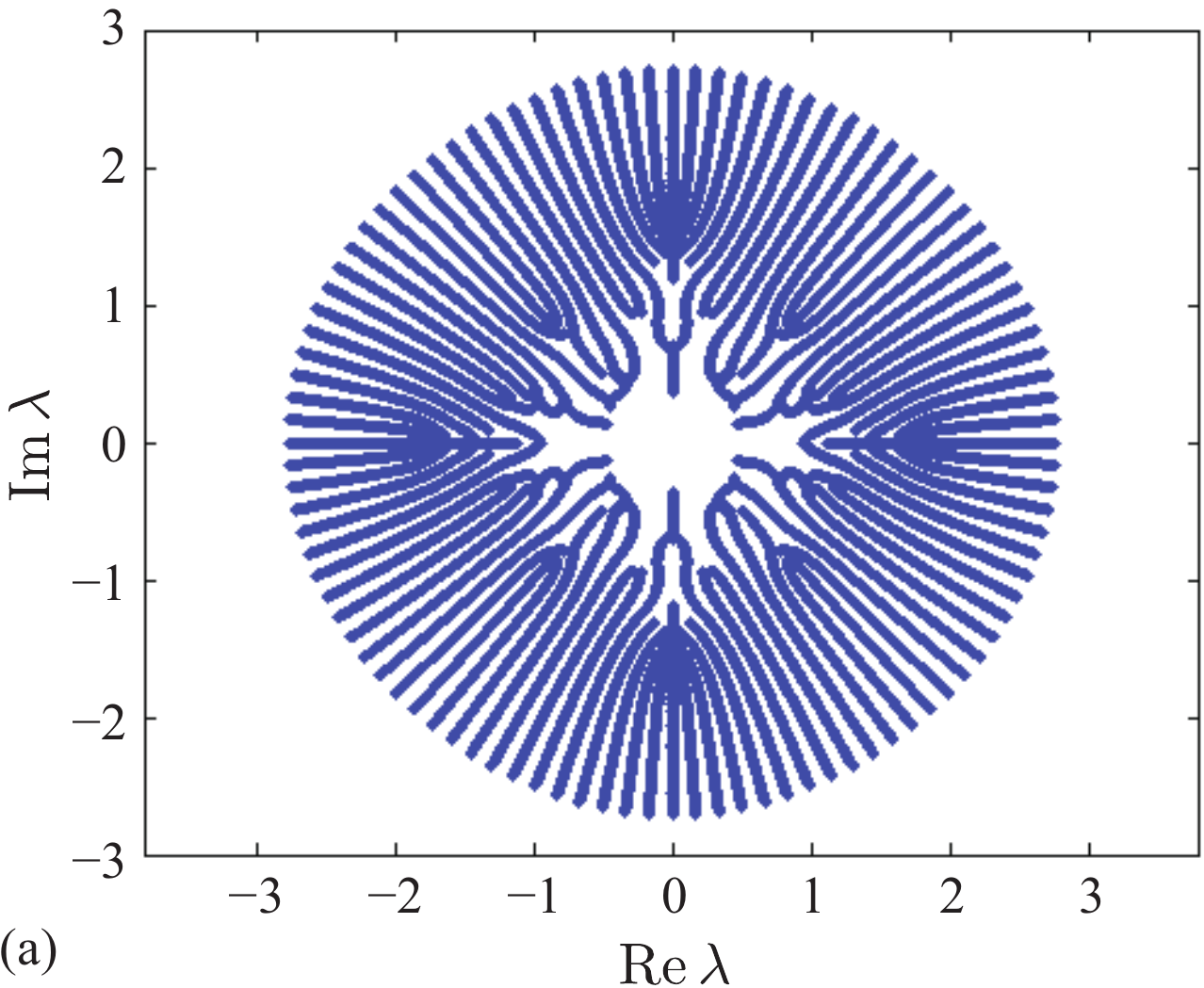}

\vspace{\baselineskip}
\includegraphics[width=.43\textwidth]{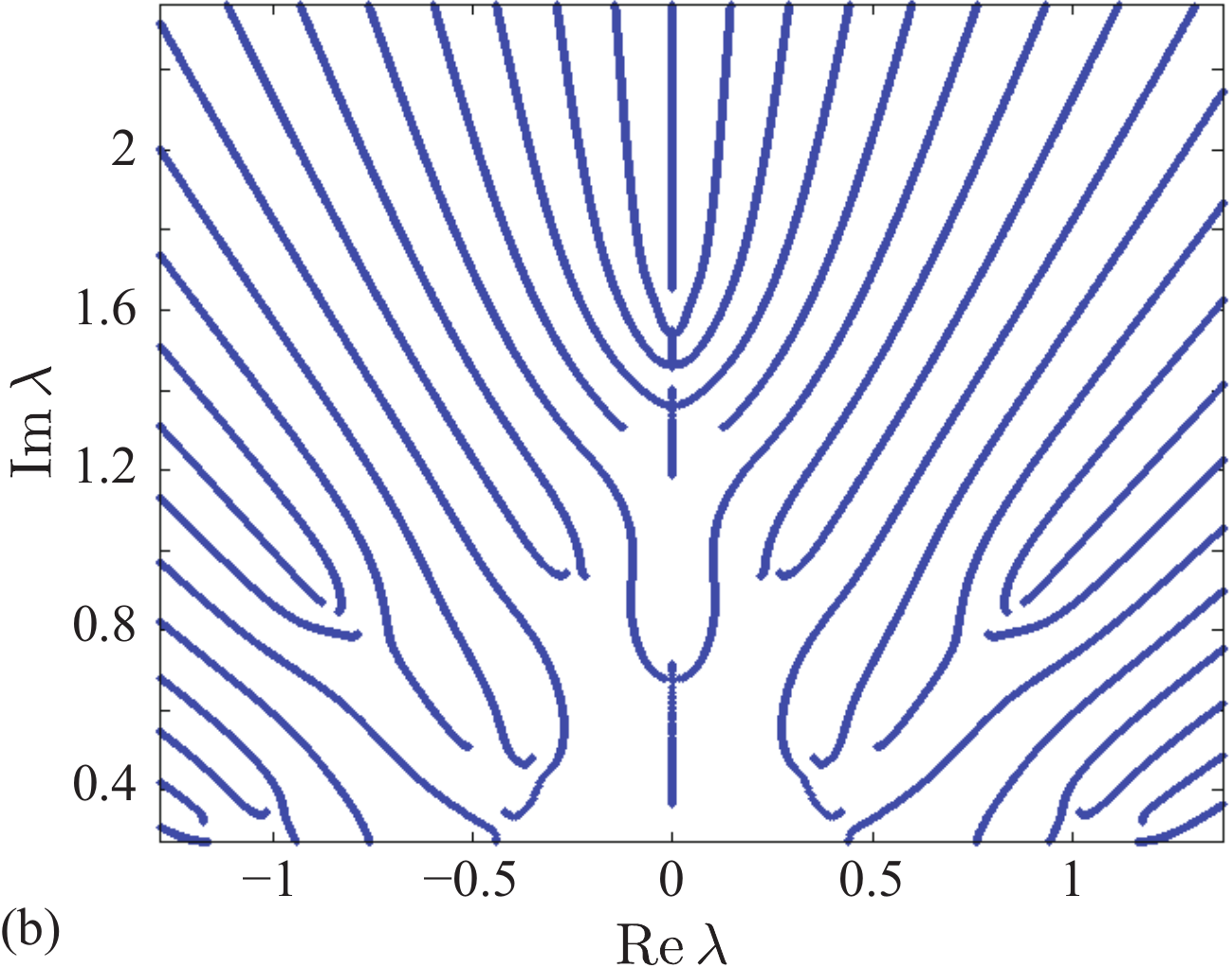}
\caption {Plot of the eigenvalues of a particular $N=100$ matrix, where $g$ is varied continuously from 1 (corresponding to the outer circle) down to 0. These eigenvalues stop changing with $g$ when their eigenfunctions localize.
(b) is an enlarged view of a part of (a).} \label {flow}
\end{figure}

Despite the deceptive simplicity of the model, it exhibits surprisingly rich behavior both in terms of the eigenvalue spectrum and in terms of the localization properties of the eigenvectors. Figure~\ref{flow} shows the trajectories of eigenvalues for a particular instance $N=100$, and for a value of $g$ decreasing from one down to zero. The eigenvalues ``flow'' in the complex plane, until their motion ultimately ceases once the corresponding eigenvectors become localized. For large values of $g$, we used perturbation theory to show that the eigenvectors are approximately plane waves (up to a similarity transformation) and that the eigenvalues form a circle (or an ellipse, more generally) in the complex plane. As $g$ decreases, eigenvalues move in the complex plane until they localize, after which ``spectral rigidity'' will take over and the motion of the localized eigenvalue stops. The final positions of the eigenvalues for $g=0$, when this game of ``musical chairs'' has ended, showed a remarkably intricate, fractal-like pattern~\cite{zee1}. For any intermediate value of $g$, the spectrum will show a pronounced ``hole'' or gap surrounding the origin, with the eigenvalues which will ultimately end inside the hole lying on its boundary, and with localized states outside it.

%

The spectra of conventional, highly-connected random matrices for large $N$ can be grouped into universality classes, such as those of the Gaussian orthogonal ensemble and the Gaussian unitary ensemble, and those obeying the Circular Law~\cite{mehta2004random}. It is natural to ask about the universality of the spectra and eigenfunctions of the one-dimensional sparse random matrices studied here. Because of its beautiful fractal-like spectrum, we have focused here on directed localization in the bimodal non-Hermitian random hopping model of Feinberg and Zee~\cite{zee1}. However, many of our conclusions also apply to the more general model defined by Eqs.~\eqref{GrindEQ__4_} and~\eqref{GrindEQ__5_}. For example, the symmetries under reflections across the real and imaginary axes and under $90^\circ$ rotations in the complex plane discussed in Sec.~\ref{symmetries} are preserved for arbitrary $u$ when $f=1/2$. As discussed in Sec.~\ref{vanish}, there is always a divergent localization length at the origin in this model. As summarized in Appendix B, when $f=1/2$, approximately equal numbers of the eigenvalues ($\sim$40-70\% total) condense onto the real and imaginary axes when $g=0$, as $u$ varies from a bimodal distribution ($u=1$) to a symmetric double box distribution ($0<u< 1$) to a symmetric single box distribution ($u=0$). As $f$ moves away from $1/2$, we expect that the spectrum becomes more elliptical, consistent with the eigenvalue spectrum derived in the large $g$ limit in Eq.~\eqref{pert2}. Another aspect of universality, that connected with Dale's law in neuroscience, was addressed in Sec.~\ref{dale_sect}.

\begin{figure}{t!}
\includegraphics[width=0.29\textwidth]{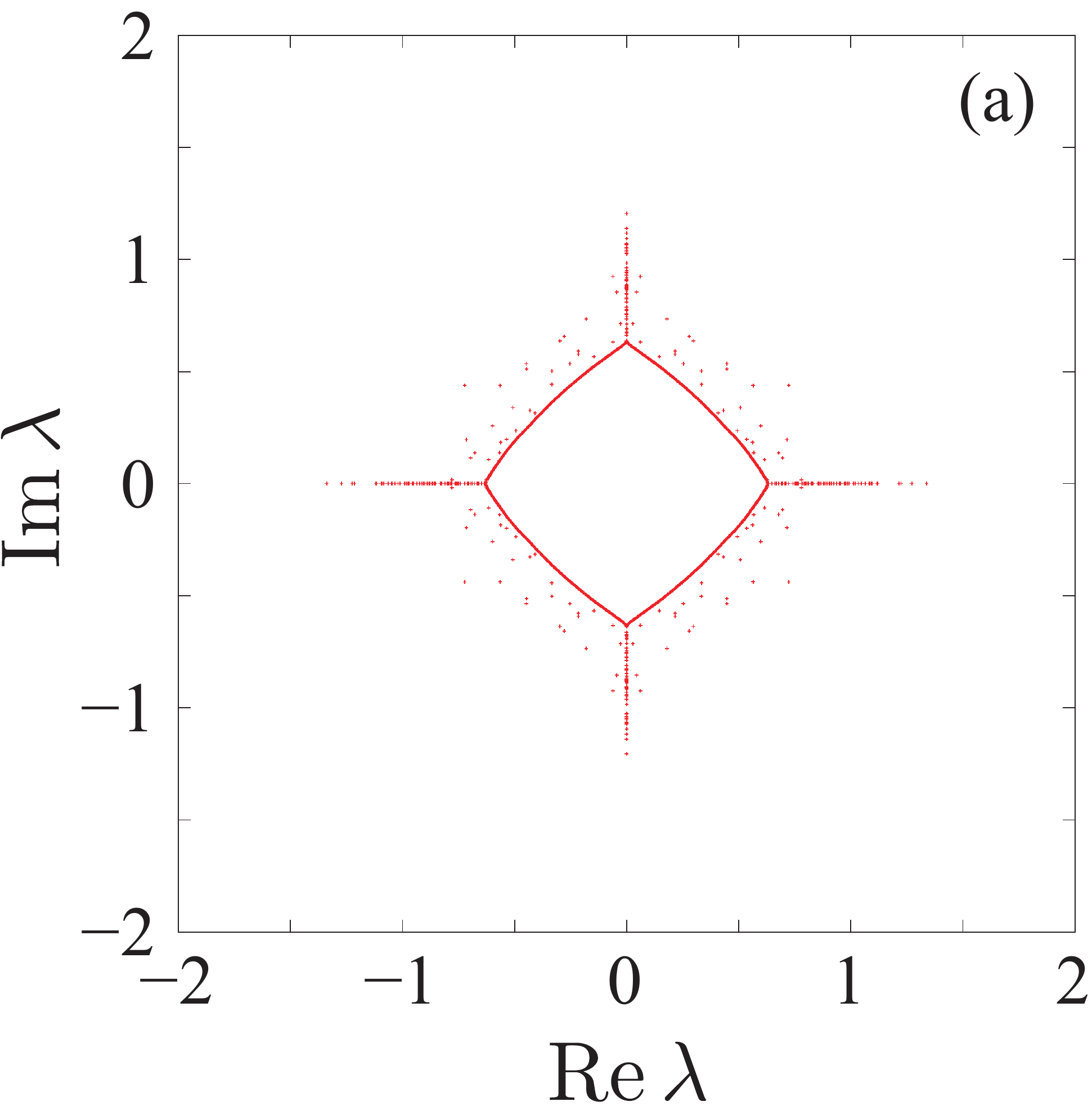}

\vspace{\baselineskip}
\includegraphics[width=0.41\textwidth]{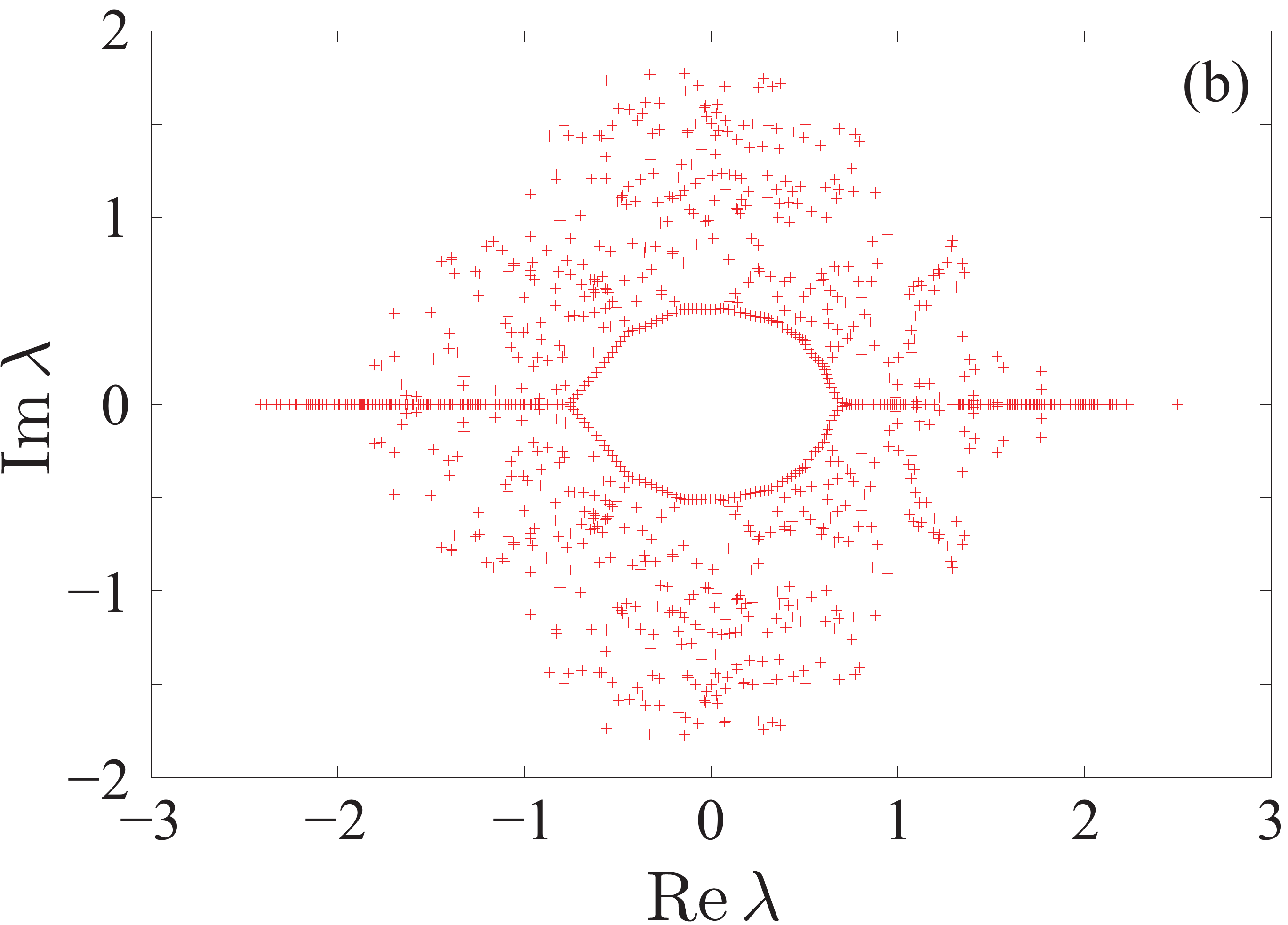}
\caption{(a)  Single box ($u = 0$) spectrum with $N=1000$ and $g=0.5$.  A macroscopic fraction of eigenvalues still condensed onto the real and imaginary axes (compare Fig.~\ref{dos}), but the remaining eigenvalues now have a diamond-like boundary.   A diamond-shaped gap now replaces the octagonal hole in Fig.~\ref{dos}(b).   (b)  Spectrum for the bimodal model with symmetrical diagonal randomness with elements $d_j$ chosen from a uniformly from the interval $[-1,1]$ , with $N = 1000$ and $g=0.1$.  Although only  the real axis now has a macroscopic fraction  of the eigenvalues,  a lens-shaped gap replaces the octagonal hole in Fig.~\ref{dos}(b).}
\label{diag_fig}
\end{figure}

The gauge transformation argument leading to Eq.~\eqref{GrindEQ__8_} is quite general. Provided the localization length increases monotonically at the origin, it predicts for 1d rings a gap or hole bounded by a rim of extended states in the spectrum for $g > g_{c_1}$.  Because the localization length diverges at the origin for the model defined by Eqs.~\eqref{GrindEQ__4_} and~\eqref{GrindEQ__5_}, we expect that $g_{c_1}=0$ in this case. When diagonal disorder is present, the localization length for $g=0$ remains finite even at the origin and now $g_{c_1}>0$~\cite{molinari}. To illustrate the universal nature of the gap, Fig.~\ref{diag_fig} shows a single box ($u=0$) spectrum with $N=1000$ and $g=0.5$ and a spectrum for the bimodal model with symmetrical diagonal randomness with elements $d_j$ chosen from a uniform distribution with support $[-1,1]$, with $N=1000$ and $g=0.1$. Although the single box spectrum in
Fig.~\ref{diag_fig}(a) no longer has the fractal-like eigenvalue spectrum shown in Fig.~\ref{dos}, a diamond-shape gap centered on the origin with an enhanced density of states is clearly present. In Fig.~\ref{diag_fig}(b) , we see that diagonal randomness added to the bimodal model destroys the symmetry under $90^\circ$ rotations by removing the eigenvalue condensation onto the imaginary axis. Nevertheless, a hole in the spectrum with an enhanced density of states on its rim survives the imposition of diagonal randomness for this value of $g =0.1>g_{c_1}>0$. The large $g$ perturbation theory of Sec.~\ref{perturb_sect} can be used to show that all states are delocalized (being plane-wave like) as $g \to \infty$ for a wide class of models, including those with diagonal randomness. Hence, we expect that there exists another critical value $g_{c_2}$, such that for $g>g_{c_2}$ all states are delocalized. Localized eigenfunctions in neuroscience could be helpful for avoiding crosstalk between different neural computation centers, and the extended states on the rim of the hole when $g > 0$ might be used to transmit information over longer distances.

Although we focus here on applications to sparse neural networks, similar non-Hermitian random matrix problems arise when random ecological networks~\cite{may1971, may1972, mccoy2011} are adapted to allow for spatial structure, with predator and prey species are localized to an array of lattice sites, but allowed to interact with their neighbors.   For example, a site dominated by foxes would have an inhibitory effect on neighboring sites occupied by rabbits, whereas rabbits would have an excitatory effect on nearby foxes.  Random excitatory and inhibitory connections in one dimension could also be studied in chains of artificial cells with spatially coupled gene expression patterns~\cite{barziv}.

\section {Acknowledgments}

We would like to thank F.~Dyson, J.~Hertz, Y.~Lue, V.~Murthy, T.~Rogers and H.~Sompolinsky for useful discussions.
Work by NH was supported by the JSPS KAKENHI Grants 15K05200, 15K05207, and 26400409.
Work by DRN was supported by the NSF, through grant DMR13063667 and through the Harvard Materials Research Science and Engineering via grant DMR1420570.

\newpage
\appendix

%

\section{Spectrum of the Hermitian random-hopping model}
\label{hermitian}

It is interesting to contrast our model of non-Hermitian localization with its Hermitian analogue, which also has a diverging localization length at the origin and a connection between the density-of-states and the inverse localization length. The Hermitian random hopping model we consider is a reformulation of Eq.~\eqref{GrindEQ__1_}
\begin{align}
\boldsymbol H_\mathrm{Herm}=-\frac{1}{2}\sum_{j=1}^N t_j\left(|j+1\rangle\langle j|+|j\rangle\langle j+1|\right),
\end{align}
where $\{t_j\}$ is a set of mutually independent random variables taking the values in the range $[1-\Delta,1+\Delta]$ with $0\leq\Delta<1$.
Although this is a standard one-dimensional version of the Anderson model~\cite{anderson,ziman}, dominated by localized eigenstates, it is well established~\cite{Theodorou76,Eggarter78} that the state at $\lambda=0$ is \emph{delocalized} with both the localization length and the density-of-states diverging as $|\lambda| \to 0$.

\begin{figure}[h!]
\includegraphics[width=0.38\textwidth]{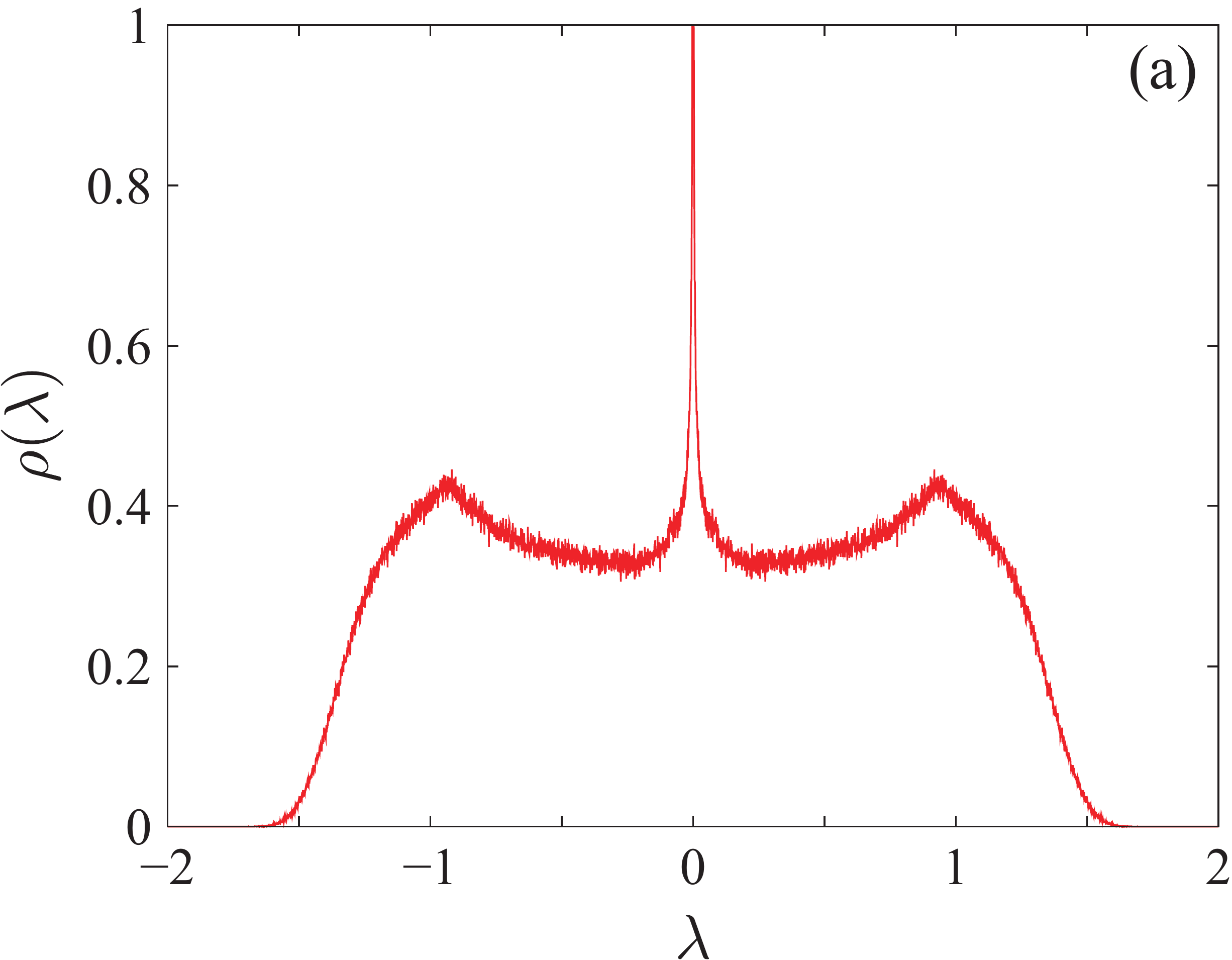}

\vspace{\baselineskip}
\includegraphics[width=0.38\textwidth]{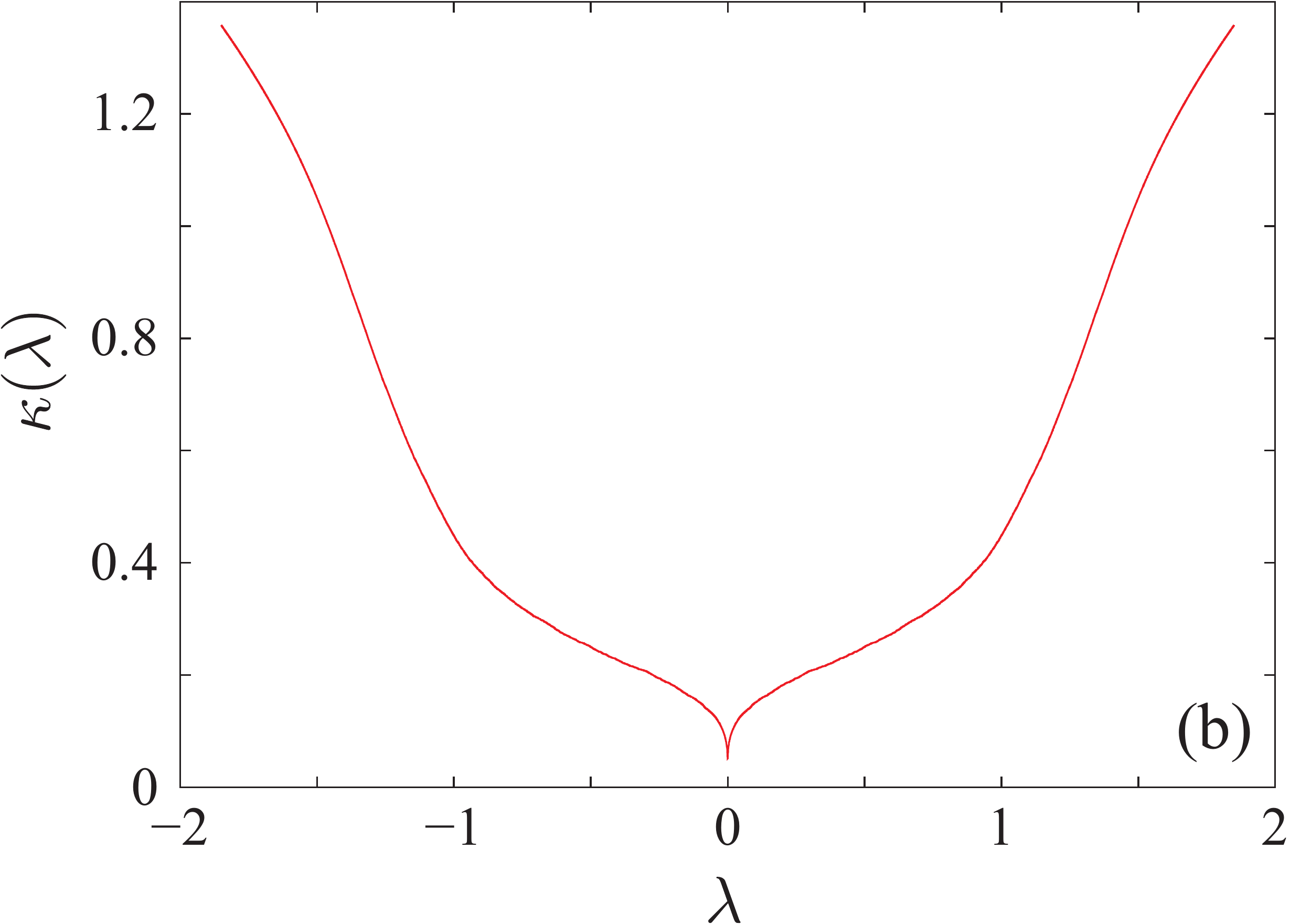}
\caption{(a) The density-of-states $\rho(\lambda)$ and (b) the inverse localization length $\kappa(\lambda)$ for a Hermitian ring of length 1000 with hopping randomness.
We computed the former from the histogram of the eigenvalues that we obtained from exact diagonalization of 5000 random samples, while the latter from a new algorithm~\cite{HatanoFeinberg15} exploiting the Chebyshev-polynomial expansion~\cite{Silver94,Silver96,Silver97} applied to 100 random samples.}
\label{app-h1fig1}
\end{figure}

Figure~\ref{app-h1fig1} illustrates the density-of-states $\rho(\lambda)$ and the inverse localization length $\kappa(\lambda)$ for $\Delta=0.85$.
We numerically confirmed that the Hermitian version of the Thouless formula connects these quantities~\cite{thouless}:
\begin{align}
 \kappa(\lambda) = P \int_{-\infty}^\infty d\lambda' \rho(\lambda') \log|\lambda-\lambda'|,
\end{align}
where $P$ denotes the principal part. We can indeed see evidence for singularities around $\lambda=0$ in both figures. These singularities are expected to take the forms~\cite{ziman}
\begin{align}
 \rho(\lambda) &\sim \frac{1}{|\lambda|\log^3 (1/|\lambda|)};\\
  \kappa(\lambda)&\sim 1/\log(1/|\lambda|).
\end{align}

\section{Density-of-states on the real and imaginary axes for $f=1/2$ and arbitrary $u$}
\label{app-h2}

In this Appendix we study the density of eigenstates that have condensed on the real and imaginary axes for the model defined by Eqs.~\eqref{GrindEQ__4_} and~\eqref{GrindEQ__5_} with $f=1/2$ and $g=0$ as a function of $u$. For numerical purposes, we here define the states to lie on the real and imaginary axes provided
\begin{align}
\left|\mathop{\mathrm{Im}}\lambda_n\right|<10^{-8},
\\
\left|\mathop{\mathrm{Re}}\lambda_n\right|<10^{-8}.
\end{align}
We list the fraction of the eigenvalues that satisfy these conditions in Table~\ref{app-h2tab1}, where the zero eigenvalues are the states that satisfy both criteria. As discussed in Sec.~\ref{loc_sect}, these states play an important role in determining how $\kappa(x,y)$ vanishes near the origin.
\begin{figure*}[t]
\includegraphics[width=0.75\textwidth]{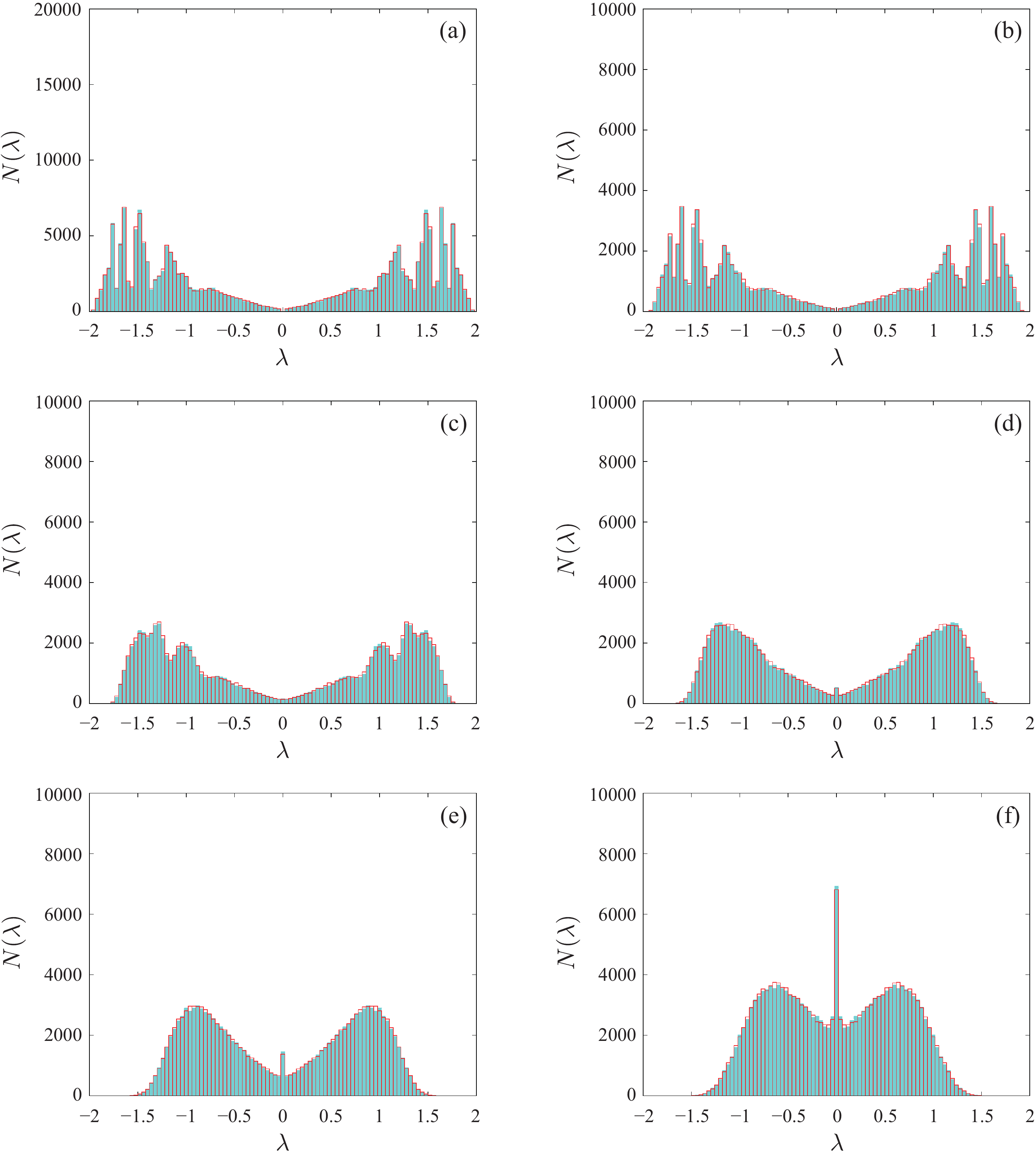}
\caption{The density-of-states for $f=1/2$, $g=0$ on the real axis (filled light blue bars) and on the imaginary axis (open red bars) plotted as histograms $N(\lambda)$:
(a) binomial distribution ($u=1$); (b) $u=0.95$; (c) $u=0.75$; (d) $u=0.5$; (e) $u=0.25$; (f) one-box distribution ($u=0$).
The system size is 1000 for all data and the number of samples is 500 for all data except for (a), where it is 1000.}
\label{app-h2fig1}
\end{figure*}

\begin{table}
\caption{The fraction of the states on the real and imaginary axes as well as of the states with the zero eigenvalue.
The data in the row ``binomial'' are for the binomial distribution $\pm 1$ of 1000 samples of length 1000, the data in the next four rows are for the two-box distribution of 500 samples of length 1000, and the data in the last row are for the one-box distribution $[-1,1]$ of 500 samples of length 1000.}
\label{app-h2tab1}
\vspace{\baselineskip}
\centering
\begin{tabular}{lrrrrr}
\hline
distribution
& \multicolumn{1}{l}{on} & \multicolumn{1}{l}{on} & \multicolumn{1}{l}{zero} \\
& \multicolumn{1}{l}{real axis} & \multicolumn{1}{l}{imaginary axis} & \multicolumn{1}{l}{eigenvalues} \\
\hline\hline
binomial ($u=1$)
& 19.9\% & 19.9\% & 0 \\
\hline
$u=0.95$ 
& 19.8\% & 20.0\% & 0 \\
\hline
$u=0.75$ 
& 20.4\% & 20.6\% & 0 \\
\hline
$u=0.5$ 
& 21.8\% & 21.9\% & 0 \\
\hline
$u=0.25$ 
& 24.7\% & 24.8\% & \; 66 (0.013\%) \\
\hline
one box (u=0) 
& 33.7\% & 33.8\% & 792 (0.16\%) \\
\hline
\end{tabular}
\end{table}

In all cases, the density-of-states (see Fig.~\ref{app-h2fig1}) is statistically the same on the real and imaginary axes.
The zero eigenvalues are absent until the two-box distribution (Eq.~\eqref{GrindEQ__5_}, $0<u<1$) becomes close to the one-box distribution ($u \lesssim 1$).
The zero-eigenvalue states would be extended if they existed for the binomial distribution, as is shown in Sec.~\ref{vanish}.
The density-of-states looks noisy for the binomial distribution;
this may reflect the fractality of the spectrum.
However, it becomes smooth for $u\leq 0.5$ and at the same time develops a peak around $\lambda=0$.

\section{Perturbation theory of the sign-random tight-binding chain}
\label{second_order}

We summarize here second-order perturbation theory applied to our model with $f=1/2$, $u=1$, for large $g$.
Upon adopting the similarity transformation of Eq.~\eqref{eq-h20},
\begin{align}\label{eq-app-h3-10}
\boldsymbol T_{jj}=\prod_{k=1}^{j-1}\frac{1}{b_k}
\end{align}
we can bring the tridiagonal hopping matrix
\begin{align}\label{eq-app-h3-20}
\boldsymbol M(g)=\begin{pmatrix}
 & s^+_1e^g &&&& s^-_N e^{-g} \\
s^-_1e^{-g} &  & s^+_2e^g &&& \\
& s^-_2e^{-g} &  & && \\
& &\ddots & &  \ddots   & \\
&&&  &  & s^+_{N-1}e^g \\
s^+_N e^{g}&&&& s^-_{N-1}e^{-g} &
\end{pmatrix},
\end{align}
into the form
\begin{align}\label{eq-app-h3-30}
 \boldsymbol M'(g) &= T^{-1}\boldsymbol M(g)\boldsymbol T \nonumber \\=&\begin{pmatrix}
 & e^g &&&& r_N e^{-g} \\
r_1e^{-g} &  & e^g &&& \\
& r_2e^{-g} &  &  && \\
& &\ddots & & \ddots   & \\
&&& &  & e^g \\
e^g&&&& r_{N-1}e^{-g} &
\end{pmatrix},
\end{align}
where all remaining matrix elements in Eqs.~\eqref{eq-app-h3-20} and~\eqref{eq-app-h3-30} are zero,
\begin{align}\label{eq-app-h3-40}
r_j=s^+_j s^-_j, \qquad j=1,..,N,
\end{align}
and we have assumed $\prod_{j=1..N}s^+_j =1$ in order to get simplified corner matrix elements.
The elements ${r_j}$ are positive or negative random numbers if $s^+_j$ and $s^-_j$ are random and both positive and negative;
In particular, when $s^+_j$ and $s^-_j$ are $\pm 1$, we have $r_j =\pm 1$ as well.
The matrices $\boldsymbol M(g)$ and $\boldsymbol M'(g)$ are then isospectral.

We now split the matrix $\boldsymbol M'(g)$ into the matrix with elements proportional to $e^g$ and the matrix with the elements proportional to $e^{-g}$, and formulate the perturbation of the spectrum of the former with respect to the latter.
We thus set $\boldsymbol M'(g)=\boldsymbol {M_0}+\boldsymbol {M_1}$, where
\begin{align}\label{eq-app-h3-50}
\boldsymbol M_0&=e^g
\begin{pmatrix}
0 & 1 & & & & \\
& 0 & 1 & & & \\
& & \ddots & \ddots & & \\
& & & 0 & 1 &\\
& & && 0 & 1 \\
1 && &&& 0
\end{pmatrix},
\\\label{eq-app-h3-60}
\boldsymbol M_1&=e^{-g}
\begin{pmatrix}
0 & &&&& r_N\\
r_1 & 0 & &&& \\
& r_2 & 0 &&&\\
&&\ddots & \ddots &&\\
&&& r_{N-2} &0 & \\
&&& &r_{N-1} &0
\end{pmatrix}.
\end{align}

\begin{figure}
\includegraphics[width=0.32\textwidth]{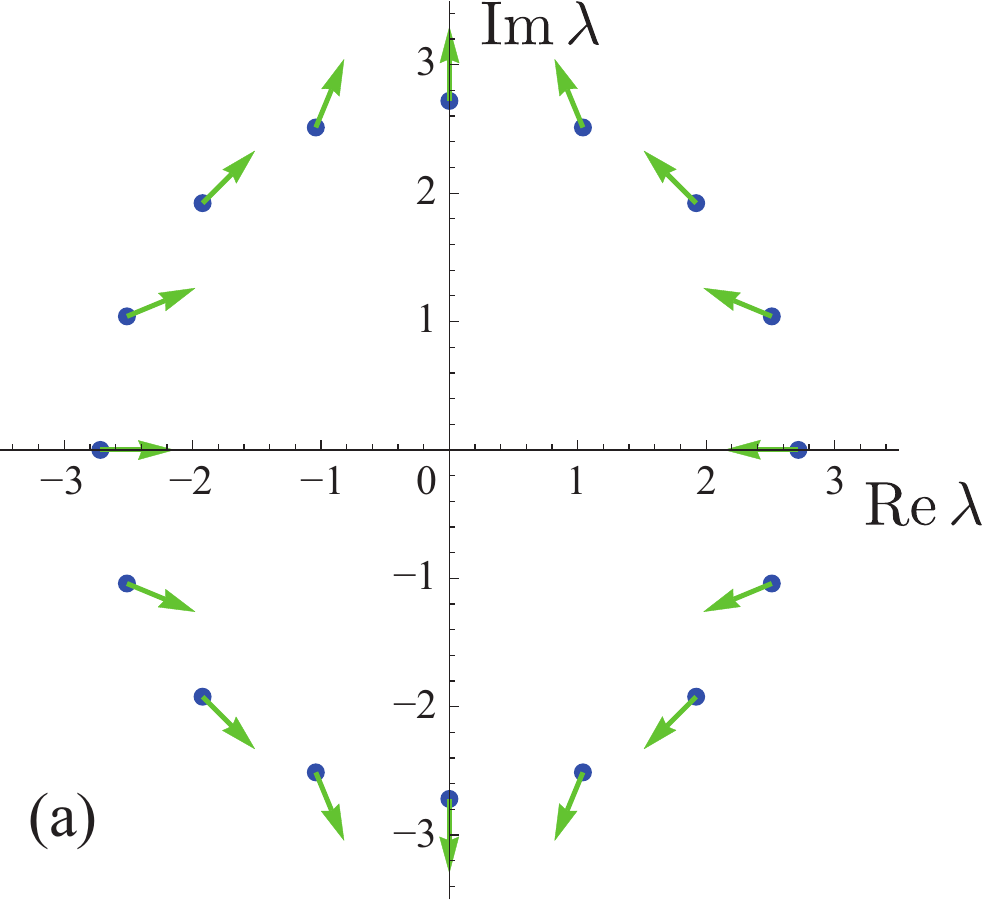}

\vspace{\baselineskip}
\includegraphics[width=0.32\textwidth]{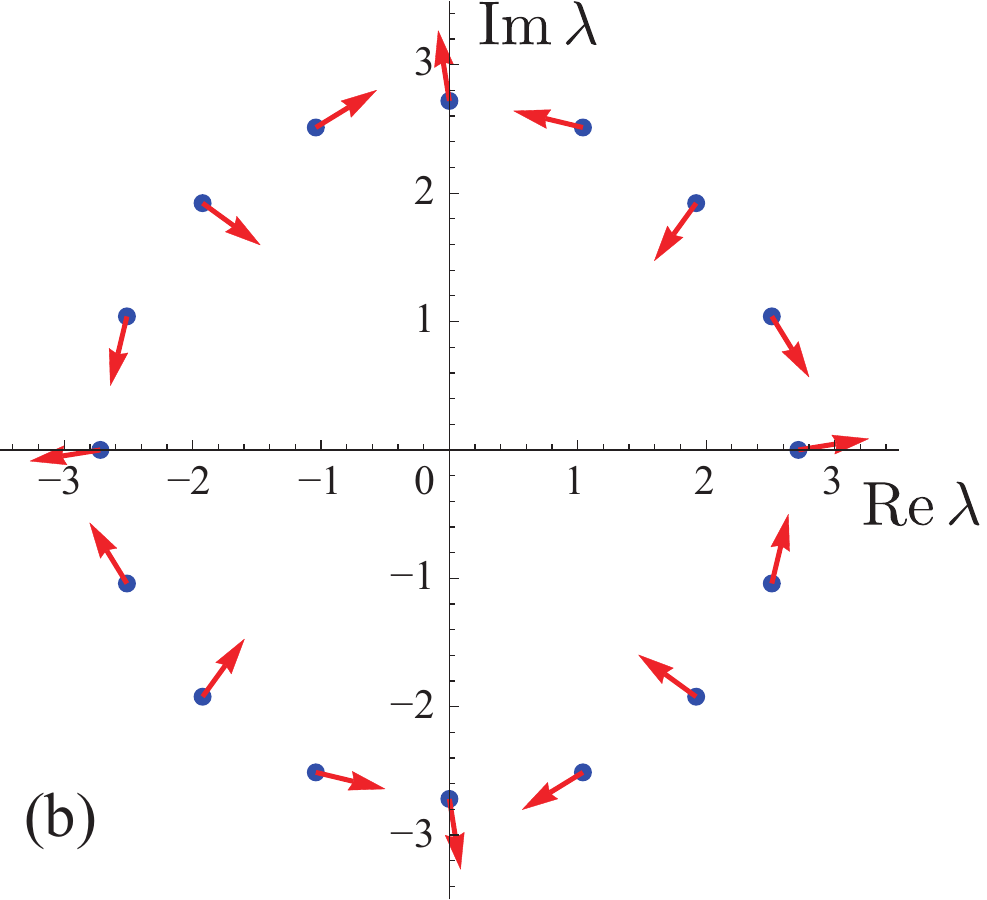}

\vspace{\baselineskip}
\includegraphics[width=0.32\textwidth]{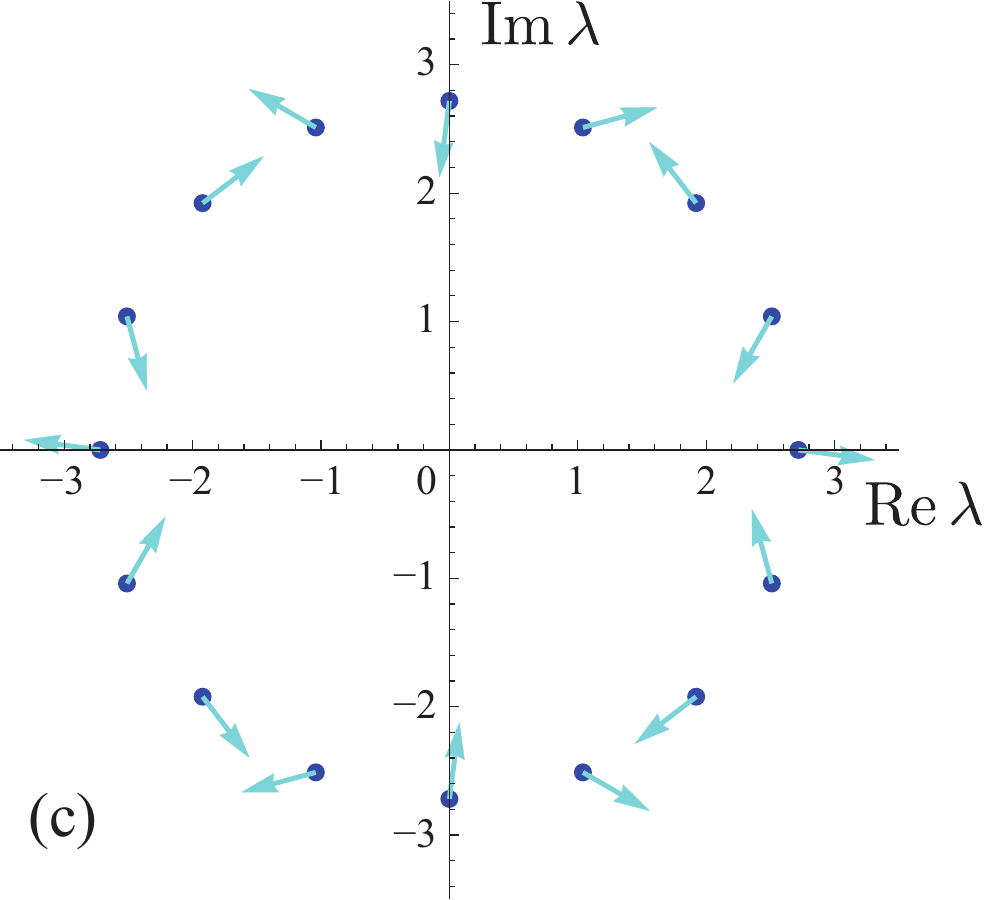}
\caption{The directions of eigenvalue perturbations with $N=16$ and $g=1$ for a random sample out of a binomial distribution.
The blue dots indicate the positions of zeroth-order eigenvalues~\eqref{eq-app-h3-80} in all panels.
The arrows in each panel indicate:
(a) the first-order perturbations~\eqref{eq-app-h3-130} magnified 50 times;
(b) the second-order perturbations~\eqref{eq-app-h3-210} magnified 20 times;
(c) The arrows indicate the third-order perturbations magnified 100 times.}
\label{app-h3-fig1}
\end{figure}

The zeroth-order eigenvalues and eigenvectors of $\boldsymbol {M_0}$ are given by
\begin{align}\label{eq-app-h3-70}
\lambda_{k_n}^{(0)}=e^{g+ik_n},
\\\label{eq-app-h3-80}
\langle x| \psi_{k_n}^{(0)}\rangle=\frac{1}{\sqrt{N}}
e^{ik_nx},
\end{align}
where
\begin{align}\label{eq-app-h3-90}
k_n&:=\frac{2\pi n}{N},\quad n=0,1,2,\cdots,N-1.
\end{align}
By setting
\begin{align}\label{eq-app-h3-100}
\xi_0=\re \lambda_{k_n}^{(0)}=e^g\cos k,
\\\label{eq-app-h3-110}
\eta_0=\im \lambda_{k_n}^{(0)}=e^g\sin k,
\end{align}
we see that the eigenvalues are equidistantly aligned on a circle of radius $e^g$ in the complex $\lambda$ plane:
\begin{align}\label{eq-app-h3-120}
{\xi_0}^2+{\eta_0}^2=e^{2g}.
\end{align}
Similarly to Sec.~\ref{perturb_sect}, we find the first-order eigenvalue perturbative shift in eigenvalues,
\begin{align}\label{eq-app-h3-130}
\lambda_{k_n}^{(1)}=
\langle \psi_{k_n}^{(0)}|A_1|\psi_{k_n}^{(0)}\rangle
=e^{-g-ik_n}\bar{r}(0),
\end{align}
where $\bar{r}(0)$ is the component at $k=0$ of
the Fourier transform of the random variable $r_x$:
\begin{align}\label{eq-app-h3-150}
\bar{r}(k)=\frac{1}{N}\sum_{x=0}^{N-1}r_{x+1}e^{ikx}.
\end{align}
We can cast it in the following way:
\begin{align}\label{eq-app-h3-160}
\xi_1&:=\re \lambda_{k_n}^{(1)}=e^{-g}\bar{r}(0)\cos k_n,
\\\label{eq-app-h3-170}
\eta_1&:=\im \lambda_{k_n}^{(1)}=-e^{-g}\bar{r}(0)\sin k_n,
\end{align}
Note that the first-order perturbation does not depend on the details of the random numbers but only on the average.
Since we use the random numbers with a symmetric probability distribution, $\bar{r}(0)$ vanishes in the limit $N\to\infty$.
For a finite value of $N$ ($N=16$), we find the movement of the eigenvalues as illustrated in Fig.~\ref{app-h3-fig1}(a).


%
%
%
The \emph{second}-order eigenvalue perturbation, obtained by similar techniques, is given by
\begin{align}\label{eq-app-h3-200}
\lambda_{k_n}^{(2)}
&:=-\sum_{\substack{m=0\\m\neq n}}^{N-1}
\frac{\langle\psi_{k_n}^{(0)}|A_1|\psi_{k_m}^{(0)}\rangle
\langle\psi_{k_m}^{(0)}|A_1|\psi_{k_n}^{(0)}\rangle}%
{\lambda_{k_m}^{(0)}-\lambda_{k_n}^{(0)}}
\\\label{eq-app-h3-210}
&=-e^{-3g}
\sum_{\substack{m=0\\m\neq n}}^{N-1}
\frac{e^{-i(k_m+k_n)}}{e^{ik_m}-e^{ik_n}}
\left|\bar{r}(k_n-k_m)\right|^2,
\end{align}
which we illustrate in Fig.~\ref{app-h3-fig1}(b).
The third-order eigenvalue perturbation is illustrated in Fig.~\ref{app-h3-fig1}(c) too.
More analyses reveal that the $k$th-order corrections generally behave as
\begin{align}\label{eq-app-h3-220}
\lambda_k^{(2)}\propto e^{-3g-3ik},\qquad
\lambda_k^{(3)}\propto e^{-5g-5ik}.
\end{align}
Although large $g$ perturbation theory is useful for capturing eigenvalue trends, it does not seem capable of determining when eigenvalues stop moving with increasing $g$; the corresponding eigenvalues remain delocalized within this approach.

%

\end{document}